\documentclass{aa}

\makeatletter
\renewcommand*\aa@pageof{, page \thepage{} of \pageref*{LastPage}}
\makeatother

\usepackage{natbib}
\bibpunct{(}{)}{;}{a}{}{,}
\usepackage{amssymb,bm}
\usepackage{url}
\usepackage{graphicx}
\usepackage{mathrsfs}
\usepackage{longtable}
\usepackage[usenames,dvipsnames]{color}
\usepackage{booktabs}
\usepackage{lscape,xcolor}

\usepackage{multirow}
\usepackage{newtxtext}
\usepackage{amsmath}
\usepackage{enumitem}
\usepackage[T1]{fontenc} 

\def\Msun{\hbox{$\rm\thinspace M_{\odot}$}}

\DeclareUnicodeCharacter{2212}{-}

\usepackage{color}
\usepackage{natbib,twoopt}
\usepackage[hyphenbreaks]{breakurl}
\usepackage[breaklinks]{hyperref}     
\bibpunct{(}{)}{;}{a}{}{,}            
\definecolor{cobalt}{rgb}{0.06, 0.2, 0.65}
\hypersetup{
  colorlinks,
  citecolor=cobalt,
  linkcolor=[rgb]{0.8, 0.2, 1.0},
  urlcolor=cobalt,
}
\makeatletter
  \newcommandtwoopt{\citeads}[3][][]{\href{http://adsabs.harvard.edu/abs/#3}%
    {\def\hyper@linkstart##1##2{}%
     \textcolor{red}\hyper@linkend\@empty\citealp[#1][#2]{#3}}}
  \newcommandtwoopt{\citepads}[3][][]{\href{http://adsabs.harvard.edu/abs/#3}%
    {\def\hyper@linkstart##1##2{}%
     \textcolor{red}\hyper@linkend\@empty\citep[#1][#2]{#3}}}
  \newcommandtwoopt{\citetads}[3][][]{\href{http://adsabs.harvard.edu/abs/#3}%
    {\def\hyper@linkstart##1##2{}%
     \textcolor{red}\hyper@linkend\@empty\citet[#1][#2]{#3}}}
  \newcommandtwoopt{\citeyearads}[3][][]%
    {\href{http://adsabs.harvard.edu/abs/#3}
    {\def\hyper@linkstart##1##2{}%
     \textcolor{red}\hyper@linkend\@empty\citeyear[#1][#2]{#3}}}
\makeatother

\begin{document}

 \title{
 Evolution of the infrared luminosity function and its corresponding dust-obscured star formation rate density out to $z\sim 6$
 }
 
\titlerunning{Infrared luminosity function out to $z\sim 6$}
\authorrunning{Koprowski et al.}

\author{M.~P.~Koprowski\inst{\ref{inst:tor}}
\and
J.~V.~Wijesekera\inst{\ref{inst:tor}}
\and
J.~S.~Dunlop\inst{\ref{inst:roe}}
\and
K.~Lisiecki\inst{\ref{inst:tor},\ref{inst:ncbj}}
\and
D.~J.~McLeod\inst{\ref{inst:roe}}
\and
R.~J.~McLure\inst{\ref{inst:roe}}
\and
M.~J.~Micha{\l}owski\inst{\ref{inst:poz},\ref{inst:roe}}
\and
M.~Solar\inst{\ref{inst:poz}}
        }

\institute{
Institute of Astronomy, Faculty of Physics, Astronomy and Informatics, Nicolaus Copernicus University, Grudzi\c{a}dzka 5, 87-100 Toru\'{n}, Poland, {\tt drelkopi@gmail.com}\label{inst:tor}
\and
Institute for Astronomy, University of Edinburgh, Royal Observatory, Edinburgh EH9 3HJ, UK \label{inst:roe}
\and
Astronomical Observatory Institute, Faculty of Physics and Astronomy, Adam Mickiewicz University, ul.~S{\l}oneczna 36, 60-286 Pozna{\'n}, Poland \label{inst:poz}
\and
National Centre for Nuclear Research, Pasteura 7, 093, Warsaw, Poland \label{inst:ncbj}
}
\abstract{We present a new determination of the evolving far-infrared (FIR) galaxy luminosity function (LF) and the resulting inferred evolution of dust-obscured star-formation rate density ($\rho_{\rm SFR}$) out to redshift $z\sim 6$. To establish the evolving comoving number density of FIR-bright objects, we made use of AS2UDS, a high-resolution ALMA follow-up study of the JCMT SCUBA-2 Cosmology Legacy Survey (S2CLS) submilliter imaging in the UKIDSS UDS survey field. In order to estimate the contributions of faint and low-mass sources, we implemented a method in which the faint-end of the IR LF is inferred by stacking (in stellar mass and redshift bins) the optical and near-infrared samples of star-forming galaxies into the appropriate FIR Herschel and submillimeter JCMT maps. Using this information we determined the faint-end slope of the FIR LF in two intermediate redshift bins (where it can be robustly established) and then adopted this result at all other redshifts. The evolution of the characteristic luminosity of the galaxy FIR LF, $L_\star$, is found to increase monotonically with redshift, evolving as $L_\star \propto z^{1.38\pm 0.07}$, while the characteristic number density, $\Phi_\star$, is well fit by a double power-law function; it is constant at $z < 2.24$ and declines as $z^{−4.95\pm 0.73}$ at higher redshifts. We then calculated the evolution of the corresponding dust-obscured star-formation rate density and compared it with the results from a number of recent studies in the literature. Our analysis confirms that dust-obscured star formation activity dominates $\rho_{\rm SFR}$ at cosmic noon but then becomes progressively less important with increasing redshift. While dusty star-forming galaxies are still found out to the highest redshifts explored here, UV-visible star formation dominates at $z > 4$, and dust-obscured activity contributes less than $25\%$ to the star formation rate density by $z\sim 6$.}


\maketitle

\section{Introduction} \label{sec:intro}

One of the main goals of modern extragalactic astronomy is to describe the evolution of star formation (SF) throughout cosmic history. The most direct way of achieving this goal is through the calibration of various relationships that characterize how SF in galaxies evolves with time. The most important examples of such relationships are the ultraviolet (UV) and infrared (IR) luminosity function (LF; e.g., \citealt{Gruppioni_2013, Koprowski_2017, Wang_2019, Bouwens_2021, Fujimoto_2024}) and the star formation rate density ($\rho_{\rm SFR}$; e.g., \citealt{Madau_2014, Dunlop_2017, Traina_2024, Liu_2025, Barrufet_2025}).

Since the discovery of the cosmic infrared background (\citealt{Puget_1996, Hauser_1998}), the IR contribution to the total $\rho_{\rm SFR}$, determined through the integration of the corresponding luminosity function, has been the primary focus of numerous studies (e.g., \citealt{Magnelli_2013, Gruppioni_2013, Madau_2014, Rowan_2016, Koprowski_2017, Dunlop_2017, Wang_2019, Gruppioni_2020, Traina_2024, Magnelli_2024, Fujimoto_2024, Sun_2025, Liu_2025}). While most of the results seem to be consistent out to $z\sim 2-3$, where the SF activity of the Universe reaches its peak, at $\gtrsim 4$ the derived values disagree by over an order of magnitude.

The accurate assessment of the evolution of SF in the early Universe is essential for evaluating current galaxy evolution models. Thanks to the sensitivity of the rest-frame UV observations, the unobscured portion of the SF in galaxies has been investigated out to redshifts as high as $\sim 11$ (e.g., \citealt{Bouwens_2014, Laporte_2016, Harikane_2024, McLeod_2024, Donnan_2024}). Due to the time required for the formation of the interstellar dust, the unobscured SF is thought to dominate the total budget at such high redshifts. Since it is now well known that by cosmic noon ($z\sim 2-3$) the vast majority of stellar emission in the UV gets absorbed by interstellar dust and reemitted in the IR, we expect $\rho_{\rm SFR}$ to transition from  being UV dominated to IR dominated. However, because of the observational limitations in the infrared (see \citealt{Casey_2014} for details), the exact redshift of this transition is yet to be determined.

The contribution of the dust-enshrouded stellar emission to the total SF budget in the early Universe has primarily been assessed through the most extreme IR sources, known as submillimeter galaxies (e.g., \citealt{Smail_1997, Hughes_1998}) or more broadly as dusty star-forming galaxies (DSFGs; \citealt{Casey_2014}), which have very large IR luminosities ($L_{\rm IR}\gtrsim 10^{12}\,{\rm L_\odot}$; e.g., \citealt{Chapman_2005, Michalowski_2017}) and extremely high SF rates (${\rm SFR}\gtrsim 100\,{\rm M_\odot\,yr^{-1}}$; e.g., \citealt{Swinbank_2014, Koprowski_2016}). Observations conducted using NASA's {\it Spitzer Space Telescope} and ESA’s {\it Herschel Space Observatory} enabled the determination of the infrared luminosity functions out to $z\simeq 2-4$ (e.g., \citealt{lefloch_2005, Caputi_2007, Rodighiero_2010, Patel_2013, Magnelli_2013, Gruppioni_2013, Rowan_2016, Wang_2019}). However, only the most extreme DSFGs have been detected at $z\gtrsim 2$, which necessitated significant extrapolations to account for fainter objects, leading to uncertain and often inconsistent results. 

\citet{Koprowski_2017} utilized a sample of the James Clerk Maxwell Telescope (JCMT) 850-${\rm \mu m}$-selected sample to construct the IR luminosity function out to $z\sim 4$. While the submillimeter wavelengths benefit from the negative $K$-correction, allowing for the observation of IR galaxies out to very high redshifts, surprisingly low numbers of sources were detected at $z\gtrsim 2$. Similarly, due to the very limited areas observed, the high-resolution ALMA blank-sky surveys conducted in recent years also struggled to detect any sources at very high redshifts (e.g., \citealt{Walter_2016, Dunlop_2017, Bouwens_2020}). \citet{Gruppioni_2020} made use of the ALMA sources serendipitously detected as a part of the ALMA Large Program to INvestigate CII at Early Times (ALPINE) in order to derive the IR LF out to $z\sim 6$. \citet{Fujimoto_2024} reached similar redshifts utilizing the ALMA Lensing Cluster Survey (ALCS) data, with faint objects detected through lensing. Most recently, \citet{Magnelli_2024, Traina_2024}, and \citet{Liu_2025} determined the contribution of IR sources to the total $\rho_{\rm SFR}$ in the early Universe using the somewhat inhomogeneous sample of individual ALMA pointings collected as a part of the A$^3$COSMOS survey \citep{Liu_2019, Adscheid_2024}. Due to the limited sensitivity of the IR data, the corresponding faint-end slope of the IR LF reported in these works ranges between -0.2 and -1.0, with errors approaching 0.5. In addition, \citet{Dudzeviciute_2020} analyzed a high-resolution ALMA follow-up study (AS2UDS; \citealt{Stach_2019}) of the JCMT SCUBA-2 submillimeter survey of the UKIDSS UDS field \citep{Geach_2017}. As a part of their analysis, \citet{Dudzeviciute_2020} derived photometric redshifts, IR luminosities, IR luminosity functions, and the corresponding star formation rate densities down to the 870\,${\rm \mu m}$ flux limit of 3.6\,mJy, later extrapolating their results down to a flux limit of 1\,mJy using the ALMA number counts of \citet{Hatsukade_2018}. They found that the contribution of submillimeter galaxies to the total star 
formation rate density in the Universe peaks at $z\sim 3$, with a contribution of $\sim 15$\% for sources with $S_{870} > 3.6$\,mJy and $\sim 60$\% for $S_{870} > 1$\,mJy, indicating that roughly half of the $\rho_{\rm SFR}$ at that redshift comes from ultra luminous IR galaxies (ULIRGs)-luminosity sources.

The aim of this study is to add to the existing knowledge by deriving the evolution of the IR LF, including its faint-end slope, up to $z\sim 6$. In order to include the contribution from the faintest galaxies, we adopted an indirect approach, where the faint-end portion of the LF is determined through stacking of the optically selected mass-complete sample of star-forming galaxies in the far-IR (FIR) maps of {\it Herschel} and JCMT maps, with the stellar mass used as a proxy for the IR luminosity. The paper is structured as follows. We present the data in Section\,\ref{sec:data}. In Section\,\ref{sec:lf1} we explain the details behind our derivation of the IR LF faint-end slope. In Section\,\ref{sec:lf2} we describe how we used data from AS2UDS, a high-resolution ALMA follow-up study of all the SCUBA-2 sources detected within the S2CLS map of the UKIDSS UDS field \citep{Geach_2017, Stach_2019}, to construct the bright end of our IR LF. In Section\,\ref{sec:lf4} we construct the functional form of the IR LF and quantify its evolution with redshift out to $z\sim 6$, discuss the results, and compare them with recent literature. The corresponding star formation rate density is derived, discussed, and compared with other works in Section\,\ref{sec:sfrd}. We provide a summary of this work in Section\,\ref{sec:sum}. Throughout the paper we assume a flat cold dark matter cosmology with $H_0=70\,{\rm km\, s^{-1}\, Mpc^{-1}}$, $\Omega_m=0.3$, and $\Omega_\Lambda=0.7$.

\section{Data} \label{sec:data}

\begin{table*}
\tiny
\caption{Infrared luminosity comoving volume density for the faint-end of the IR LF.}\label{tab:fe}
\centering
\renewcommand{\arraystretch}{1.5}
\setlength{\tabcolsep}{4pt}
\begin{tabular}{ccccccc}
\hline
\hline
& \multicolumn{6}{c}{${\rm log}(\Phi/{\rm Mpc^{−3}\,dex^{-1}})$} \\
& $9.75\leq \mathcal{M}<10.00$ & $10.00\leq \mathcal{M}<10.25$ & $10.25\leq \mathcal{M}<10.50$ & $10.50\leq \mathcal{M}<10.75$ & $10.75\leq \mathcal{M}<11.00$ & $11.00\leq \mathcal{M}<11.25$ \\
\hline
& $(\mathcal{L}=10.91 \pm 0.29)$ & $(\mathcal{L}=11.14 \pm 0.27)$ & $(\mathcal{L}=11.35 \pm 0.25)$ & $(\mathcal{L}=11.53 \pm 0.23)$ & $(\mathcal{L}=11.67 \pm 0.23)$ &  \\
$\bar{z}=1.44$ &$-2.35^{+0.03}_{-0.03}\phantom{-}$ & $-2.45^{+0.04}_{-0.04}\phantom{-}$ & $-2.53^{+0.06}_{-0.08}\phantom{-}$ & $-2.60^{+0.08}_{-0.10}\phantom{-}$ & $-2.73^{+0.12}_{-0.18}\phantom{-}$ & -- \\
&  & $(\mathcal{L}=11.31 \pm 0.27)$ & $(\mathcal{L}=11.52 \pm 0.26)$ & $(\mathcal{L}=11.72 \pm 0.24)$ & $(\mathcal{L}=11.88 \pm 0.22)$ &  \\
$\bar{z}=1.88$ &-- & $-2.58^{+0.03}_{-0.03}\phantom{-}$ & $-2.67^{+0.05}_{-0.05}\phantom{-}$ & $-2.77^{+0.08}_{-0.09}\phantom{-}$ & $-2.88^{+0.10}_{-0.13}\phantom{-}$ & -- \\
&  &  & $(\mathcal{L}=11.72 \pm 0.26)$ & $(\mathcal{L}=11.92 \pm 0.24)$ & $(\mathcal{L}=12.11 \pm 0.23)$ & $(\mathcal{L}=12.25 \pm 0.20)$ \\
$\bar{z}=2.59$ &-- & -- & $-2.86^{+0.04}_{-0.04}\phantom{-}$ & $-3.04^{+0.06}_{-0.06}\phantom{-}$ & $-3.21^{+0.08}_{-0.10}\phantom{-}$ & $-3.47^{+0.11}_{-0.15}\phantom{-}$ \\
&  &  &  & $(\mathcal{L}=12.15 \pm 0.23)$ & $(\mathcal{L}=12.37 \pm 0.21)$ & $(\mathcal{L}=12.53 \pm 0.19)$ \\
$\bar{z}=4.14$ &-- & -- & -- & $-3.83^{+0.04}_{-0.05}\phantom{-}$ & $-4.11^{+0.06}_{-0.07}\phantom{-}$ & $-4.34^{+0.09}_{-0.11}\phantom{-}$ \\
\hline
\end{tabular}
\tablefoot{$\mathcal{M}\equiv {\rm log}(M_\ast/{\rm M_\odot})$ and $\mathcal{L}\equiv {\rm log}(L_{\rm IR}/{\rm L_\odot})$. For each stellar mass bin, the corresponding IR luminosity at a given redshift is shown in the brackets. Details are provided in Section\,\ref{sec:lf1}.}
\end{table*}

The faint end of the infrared luminosity function was established via stacking the optical/near-IR catalogs of the UKIDSS Ultra Deep Survey (UDS) and COSMOS fields (\citealt{McLeod_2021}) following the methodology outlined in \citet{Koprowski_2024}, where the stellar masses of galaxies were used as a proxy for the IR luminosity. For the purpose of extracting the stacked FIR flux densities, we used the {\it Herschel} \citep{Pilbratt_2010} Multi-tiered Extragalactic Survey (HerMES; \citealt{Oliver_2012}) and the Photodetector Array Camera and Spectrometer (PACS; \citealt{Poglitsch_2010}) Evolutionary Probe (PEP; \citealt{Lutz_2011}) data obtained with the Spectral and Photometric Imaging Receiver (SPIRE; \citealt{Griffin_2010}) and PACS instruments, where {\it Herschel} maps at 100, 160, 250, 350, and 500\,${\rm \mu m}$ were utilized. In order to constrain the Rayleigh–Jeans tail of the dust emission curve, we also include the data collected as part of the SCUBA-2 Cosmology Legacy Survey (S2CLS; \citealt{Geach_2017}).

The bright end of the IR LF (Section\,\ref{sec:lf2}) was derived using high-resolution ALMA follow-up data (AS2UDS; \citealt{Stach_2019, Dudzeviciute_2020}) of the S2CLS UKIDSS UDS field \citep{Geach_2017}. The original SCUBA-2 map encompasses an area of 0.96 square degrees, with a noise level of less than $1.3$\,mJy and a median sensitivity of $\sigma_{850} = 0.88$ mJy beam$^{-1}$, where 716 sources were detected. As explained in \citet{Geach_2017}, the SCUBA-2 source detection completeness for the UDS field was determined by the recovery rate of artificial sources injected into the jackknife maps as a function of input flux.

A detailed description of the ALMA observations and data reduction and the construction of the catalog can be found in \citet{Stach_2019}. In summary, all 716 SCUBA-2 sources were observed using ALMA Band 7 during Cycles 1, 3, 4, and 5. All the maps were tapered to 0.5\,arcsec for source detection purposes, with the resulting AS2UDS catalog consisting of 708 individual ALMA-identified submillimeter galaxies ($\sigma_{870} > 4.3$). Based on 60,000 simulated ALMA observations, the sample was found to be complete at $S_{870} \gtrsim 4$\,mJy. 

The associated photometric redshifts and FIR luminosities for all AS2UDS sources were adopted from \citet{Dudzeviciute_2020}. In short, the available photometry spans a wavelength range from the $u$-band to radio, with low-resolution {\it Herschel} SPIRE data deblended using the procedure explained in \citet{Swinbank_2014}. Photometric redshifts as well as IR luminosities were found using the updated {\sc magphys} spectral energy distribution modeling code \citep{daCunha_2015, Battisti_2019}, along with the stellar models from \citet{Bruzual_2003}, the initial mass function of \citet{Chabrier_2003}, and the dust attenuation model of \citet{Charlot_2000}. The resulting photometric redshifts were tested against the available spectroscopic data, yielding a median offset -- defined as $(z_{\rm spec}-z_{\rm phot})/(1+z_{\rm spec})$ -- of $−0.005 \pm 0.003$, with a dispersion of 0.13.

The corresponding dust temperatures were found by \citet{Dudzeviciute_2020} using {\sc magphys} and the modified blackbody curve:

\begin{equation}\label{eq:gb}
S\propto (1-e^{-\tau_{\rm rest}})\times B(\nu_{\rm rest}, T),
\end{equation}

\noindent where $B(\nu_{\rm rest}, T)$ is the Planck function and the optical depth defined is as $\tau_{\rm rest}\equiv (\nu_{\rm rest}/\nu_0)^\beta$, with the emissivity index $\beta$ set to 1.8. As explained in \citet{Dudzeviciute_2020}, the corresponding temperatures for both {\sc magphys} and the modified blackbody fits were found to be in a very good agreement, with a typical fractional difference of $(T_d^{\rm MBB}-T_d^{\rm MAGPHYS})/T_d^{\rm MBB}=-0.28\pm0.01$.

\section{Analysis and discussion}\label{sec:an}

\subsection{Infrared luminosity function} \label{sec:lf}

\subsubsection{Faint end} \label{sec:lf1}

\begin{figure}
\centering
   \includegraphics[width=9cm]{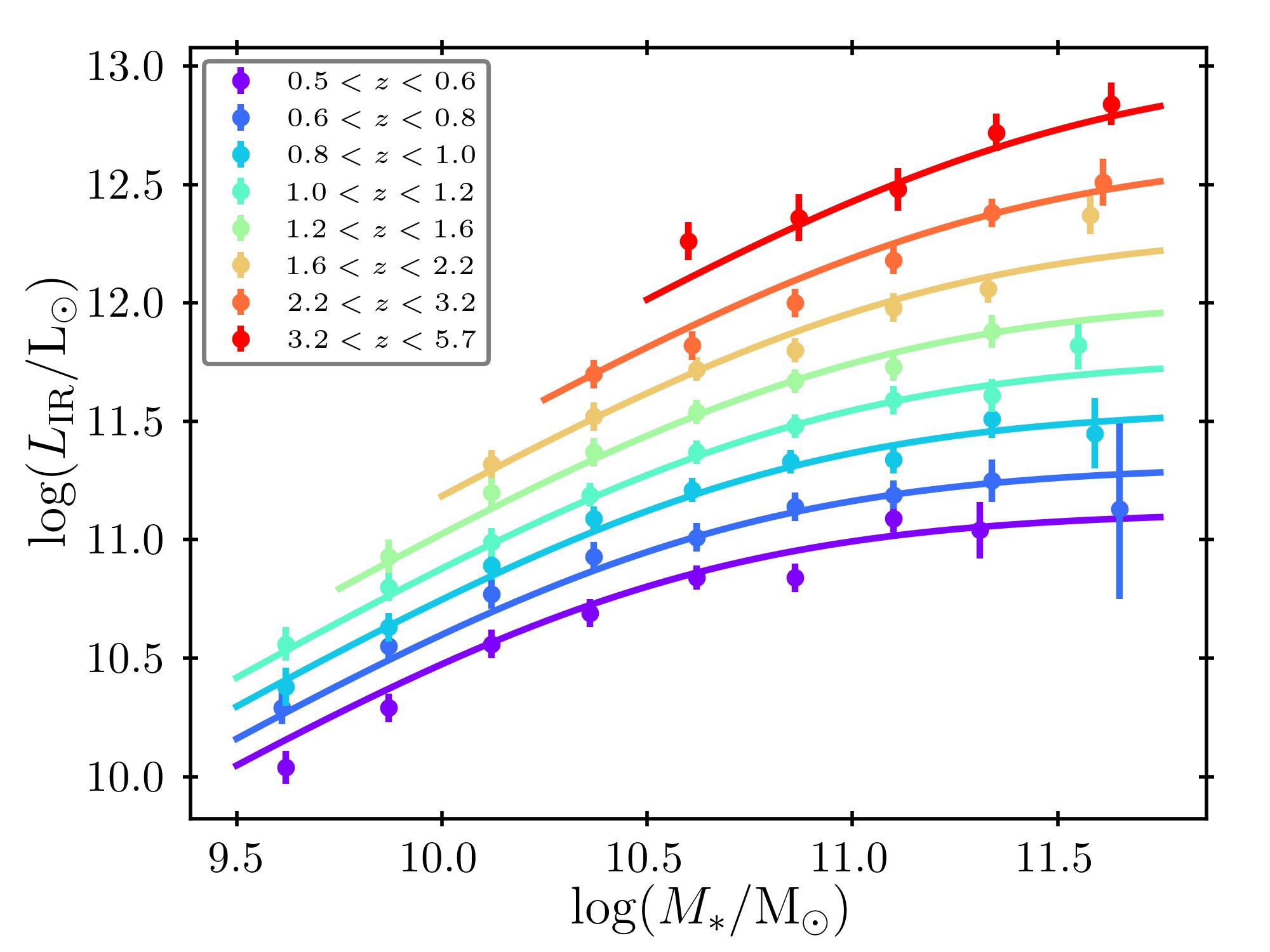}
     \caption{Infrared luminosity-stellar mass relation determined through stacking the optical/near-IR mass-complete samples of \citet{McLeod_2021} in the FIR {\it Herschel} and JCMT maps, adopted from \citet{Koprowski_2024}. The stacked $L_{\rm IR}$ values were used to determine the faint-end portion of the IR LF, as explained in Section\,\ref{sec:lf1}.}
     \label{fig:lir}
\end{figure}

The IR luminosities for the galaxies constituting the faint-end portion of the luminosity function were extracted indirectly by stacking the optical/near-IR samples in the FIR {\it Herschel} and JCMT maps and using the stellar mass as proxies. The rationale behind this approach is that the IR luminosities of the star-forming galaxies are tightly correlated with their stellar masses via the so-called main sequence (e.g., \citealt{Speagle_2014, Tomczak_2016, Daddi_2022, Popesso_2023, Koprowski_2024}). The stacking procedure was performed by \citet{Koprowski_2024}, and we adopted their results in this work (Figure\,\ref{fig:lir} with the individual $L_{\rm IR}$ values listed in Table\,\ref{tab:lir}). The redshift bins were designed to encompass the approximately 1 billion years of Universe evolution, ranging from 1 to 9 billion years after the Big Bang (with the corresponding redshift bins of [0.45, 0.6, 0.75, 1.0, 1.2, 1.6, 2.2, 3.2, 5.7]).

As we consider a sample of star-forming galaxies here, our assessment of the faint end of the IR luminosity function in effect excludes both quiescent and starburst galaxies (see \citealt{Koprowski_2024} for details). This is warranted, as passive sources are much less numerous than the active galaxies, and they exhibit significantly lower IR luminosities at a given stellar mass, rendering their contribution to the resulting LF negligible. Starbursts, on the other hand, are much more luminous in the IR, and their contribution to the LF is therefore assessed using the ALMA data, as explained in Section\,\ref{sec:lf2}.

Since the $L_{\rm IR}$-$M_\ast$ relation, similarly to the star-forming main sequence, exhibits a turnover at high masses, the functional form of \citet{Lee_2015} was adopted:

\begin{equation}\label{eq:lir}
L_{\rm IR} = L_{\rm IR}^{{max}}/(1+(M_\ast/M_0)^{-\gamma}),
\end{equation}

\noindent with the slope, $\gamma$, set to one. As explained in \citet{Koprowski_2024}, the most accurate fits to the $L_{\rm IR}$-$M_\ast$ relation are produced when the logarithm of the free parameters of Eq. \ref{eq:lir} are assumed to follow an exponential evolution with redshift, where

\begin{equation}\label{eq:pars}
\begin{split}
{\rm log}(L_{\rm IR}^{max}/{\rm M_\odot yr^{-1}}) & = a_1+a_2\times e^{-a_3z} \\
{\rm log}(M_0/{\rm M_\odot}) & = b_1+b_2\times e^{-b_3z}.
\end{split}
\end{equation}

\begin{table}
\caption{Best-fit values to parameters from Eq. \ref{eq:pars}.}\label{tab:parslir}
\centering
\begin{tabular}{cc}
\hline\hline
Parameter & Best-fit values \\
\hline
a1 & $\phantom{-1}2.97\pm 0.20$ \\
a2 & \phantom{1}$-2.62\pm 0.14$ \\
a3 & $\phantom{-1}0.59\pm 0.10$ \\
\hline
b1 & $\phantom{-}11.38\pm 0.39$ \\
b2 & \phantom{1}$-1.14\pm 0.30$ \\
b3 & $\phantom{-1}0.50\pm 0.36$ \\
\hline
\end{tabular}
\end{table}

\noindent Eq. \ref{eq:lir} was fit to the stacked data (Table\,\ref{tab:lir}) using nonlinear least square fitting. The best-fit curves are shown in Figure\,\ref{fig:lir}, with the best-fit parameters of Eq.\,\ref{eq:pars} listed in Table\,\ref{tab:parslir}.

\begin{table*}
\tiny
\caption{Infrared luminosity comoving volume density for the bright end of the IR LF.}\label{tab:be}
\centering
\renewcommand{\arraystretch}{1.5}
\begin{tabular}{ccccc}
\hline
& \multicolumn{4}{c}{${\rm log}(\Phi/{\rm Mpc^{−3}\,dex^{-1}})$} \\
& $12.50\leq \mathcal{L}<12.75$ & $12.75\leq \mathcal{L}<13.00$ & $13.00\leq \mathcal{L}<13.25$ & $13.25\leq \mathcal{L}<13.50$ \\
\hline
$\bar{z}=1.44$ & $-6.02^{+0.46}_{-{\rm inf}}\phantom{-}$ & $-6.01^{+0.34}_{-{\rm inf}}\phantom{-}$ & ... & ... \\
$\bar{z}=1.88$ & ... & $-5.61^{+0.20}_{-0.38}\phantom{-}$ & $-6.24^{+0.40}_{-{\rm inf}}\phantom{-}$ & ... \\
$\bar{z}=2.59$ & ... & $-5.07^{+0.07}_{-0.08}\phantom{-}$ & $-5.98^{+0.24}_{-0.59}\phantom{-}$ & ... \\
$\bar{z}=4.14$ & ... & ... & $-5.92^{+0.11}_{-0.15}\phantom{-}$ & $-6.82^{+0.34}_{-{\rm inf}}\phantom{-}$ \\
 \hline
\end{tabular}
\tablefoot{$\mathcal{L}\equiv {\rm log}(L_{\rm IR}/{\rm L_\odot})$. Details are provided in Section\,\ref{sec:lf2}.}
\end{table*}

Because of the lack of ALMA detections (IR LF bright end) at low redshifts (Section\,\ref{sec:lf2}), we limited our analysis to the redshift bins of [1.2, 1.6, 2.2, 3.2, 5.7]. In addition, we only considered $z$-$M_\ast$ bins with stellar masses below the main-sequence bending mass, $M_0$, at which point the $L_{\rm IR}$-$M_\ast$ relationship leaves the power-law regime ($M_0/{\rm M_\odot}\sim 10^{11.0}$ at $1.2<z<2.2$ rising to $M_0/{\rm M_\odot}\sim 10^{11.2}$ at $z>2.2$). The IR luminosity for each $z$-$M_\ast$ bin was extracted from the $L_{\rm IR}$-$M_\ast$ best-fit relation (Equations \ref{eq:lir} and \ref{eq:pars} with the best-fit parameters from Table\,\ref{tab:parslir}). In order to determine the corresponding IR luminosity function, a standard $1/V_{\rm max}$ method \citep{Schmidt_1968} was used:

\begin{equation}
\Phi(L,z)=\frac{1}{\Delta {\rm log}L}\sum_i \frac{1}{V_i}.
\end{equation}

\noindent The width of the $L_{\rm IR}$ bin in log space, $\Delta {\rm log}L$, was taken from the best-fit $L_{\rm IR}$-$M_\ast$ functional form of Equation\,\ref{eq:lir}. Given that the width of the stellar mass bin is $\Delta{\rm log}(M_\ast/{\rm M_\odot})=0.25$ and the slope, $\gamma$, in Equation\,\ref{eq:lir} was set to one, $\Delta {\rm log}L$ below the bending mass, $M_0$, is also $\simeq$0.25. The number of sources in each bin, $i$, is listed at the bottom panel of Table\,\ref{tab:lir}, and $V_i$ is the comoving volume available to the $i$th source.

The resulting values for the faint-end portion of the IR LF are listed in Table\,\ref{tab:fe} and depicted as black points in Figure\,\ref{fig:phi}. To assess the errors on the number of sources in each bin, the bootstrapping method was used. In each step, a mock catalog was constructed by drawing, at random and with replacement, a sample of sources from the original mass-complete dataset. The process was repeated 1000 times, and the uncertainties on the number of sources in each $z$-$M_\ast$ bin were taken to be the standard deviation of the resulting simulated values. The errors on the average stacked values of $L_{\rm IR}$, listed in Table\,\ref{tab:lir}, are from \citet{Koprowski_2024}. The errors on the IR LF faint-end data points are driven by the errors on the number of sources in each bin (Bottom panel of Table\,\ref{tab:lir}), and as can be seen in Figure\,\ref{fig:phi} and Table\,\ref{tab:fe}, they are significantly smaller than the errors on the $L_{\rm IR}$.

\subsubsection{Bright end} \label{sec:lf2}

The bright end of the IR LF was determined using high-resolution ALMA follow-up data of all the S2CLS UKIDSS UDS sources from AS2UDS \citep{Stach_2019, Dudzeviciute_2020}. The main advantage of the ALMA data is the high resolution of the observations, which allow for robust optical counterpart identification. Each source has two values of completeness associated with it -- one from the original S2CLS survey \citep{Geach_2017} and one from the ALMA follow-up observations \citep{Stach_2019}. To make sure the data used here are flux complete, we limited the ALMA sample to sources with fluxes $S_{870}>4$\,mJy, for which the completeness in the ALMA maps was established to be close to 100\%. Since, as stated in \citet{Stach_2019}, for SCUBA-2 sources above $S_{850} \sim 3.5$\,mJy all the single-dish flux in ALMA maps, on average, has been recovered, we can treat ALMA $S_{870}>4$\,mJy sources as a flux-complete ($>80$\% completeness in the original S2CLS SCUBA-2 UDS map of \citealt{Geach_2017}) homogeneous sample.

As discussed briefly in Section\,\ref{sec:data}, the IR luminosities were taken from \citet{Dudzeviciute_2020}, where in order to get reliable values of $L_{\rm IR}$, we limited our sample to sources with at least one {\it Herschel} SPIRE detection. We therefore required our sample to be 870${\rm \mu m}$ flux complete ($S_{870}>4$\,mJy) and have at least one SPIRE detection. The median dust temperature of the sources in the AS2UDS sample was estimated by \citet{Dudzeviciute_2020} to be $\sim 30$\,K (with a 68th percentile range of 25.7-37.3\,K). Adopting the modified blackbody curve of Equation\,\ref{eq:gb}, with the 870\,${\rm \mu m}$ flux of 4\,mJy and the dust temperature of 37.3\,K (upper 68th percentile), we arrived at the conservative IR luminosity completeness limit of ${\rm log}(L_{\rm IR}/{\rm L_\odot})=12.5$ at $z<1.6$, respectively rising to 12.75 and 13.00 at $z<3.2$ and $<5.7$ (Table\,\ref{tab:be}).

The IR luminosity function was found using the $1/V_{\rm max}$ method \citep{Schmidt_1968}, where

\begin{equation}\label{eq:lfbe}
\Phi(L,z)=\frac{1}{\Delta {\rm log}L}\sum_i \frac{1-{\rm FDR}}{w_i\times V_i},
\end{equation}

\noindent with $\Delta {\rm log}L$ being the width of the luminosity bin in log space and $V_i$ as the comoving volume available to the $i$th source, while the false detection rate (FDR) and the completeness, $w_i$, are from the original SCUBA-2 survey \citep{Geach_2017}. As explained in \citet{Geach_2017}, a fraction of low signal-to-noise ratio sources may be missed by the source detection algorithm due to them residing in the noise valleys. For the purpose of the sample completeness, the fraction of missed galaxies was therefore quantified in the form of the source completeness, $w_i$. Similarly, false sources may be identified due to statistical fluctuations expected from Gaussian noise, as quantified by FDR. In both cases, as described in \citet{Geach_2017}, the jackknife noise-only map was populated with artificial sources, after which the source extraction algorithm was applied. The ratio between the recovered and the injected numbers were then used to determine $w_i$ and FDR for each source in the parent SCUBA-2 sample of \citet{Geach_2017}, which were then adopted in our calculations of the IR LF of Equation\,\ref{eq:lfbe}. The resulting values are summarized in Table\,\ref{tab:be} and depicted as color points with gray error bars in Figure\,\ref{fig:phi}. In order to assess the errors, we conducted Monte Carlo simulations in which redshifts and infrared luminosities were randomly sampled from normal distributions, with the means and standard deviations corresponding to the catalog values and their associated errors. The $1\sigma$ uncertainties associated with the stacked IR LFs were then obtained by computing the standard deviations of the simulated data.

We note that our derived form of the bright end of the IR LF closely resembles that presented in \citet{Dudzeviciute_2020}. The only notable distinctions are that we have assumed slightly different redshift binning and included the completeness and false detection rate values of the original S2CLS sample of \citet{Geach_2017} in our $1/V_{\rm max}$ analysis. However, given that the initial SCUBA2 sample has been determined to be nearly flux complete for $S_{850}>4$\,mJy, incorporating them did not have a substantial impact on our results. The comparison with the IR LF and the $\rho_{\rm SFR}$ values found in \citet{Dudzeviciute_2020} is presented in Sections\,\ref{sec:lf4} and \ref{sec:sfrd}.

\subsubsection{IR LF functional form}\label{sec:lf4}

\begin{figure}
\centering
   \includegraphics[width=9cm]{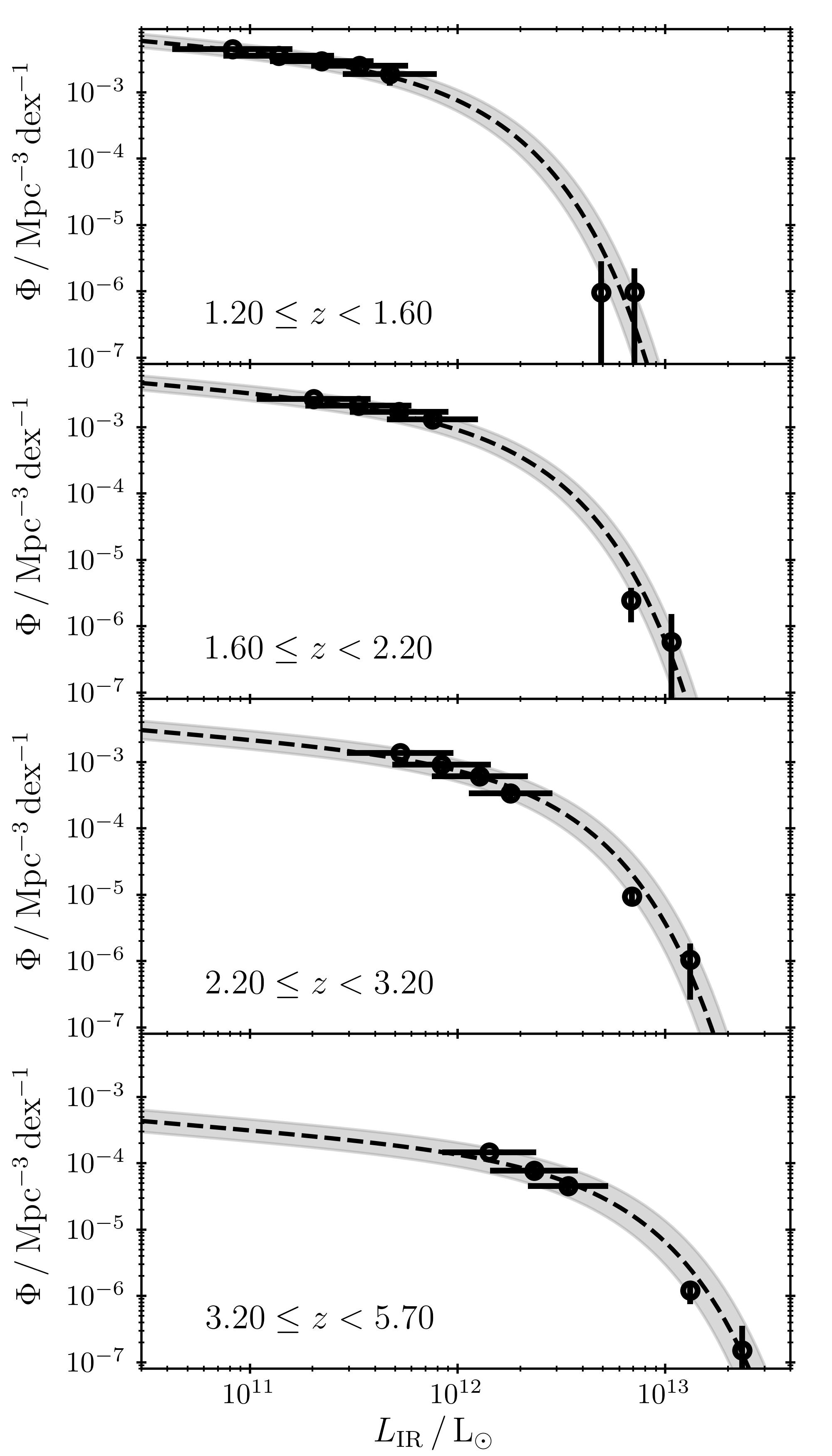}
     \caption{Infrared luminosity function found in this work at different redshift bins, as indicated in the plot. The faint-end data (Table\,\ref{tab:fe}), determined through stacking (see Section\,\ref{sec:lf1} for details), are depicted by black points with horizontal error bars. The remaining bright-end data (Table\,\ref{tab:be}) found using the ALMA follow-up data of the S2CLS UKIDSS UDS sources (AS2UDS; \citealt{Stach_2019}) are also shown (Section\,\ref{sec:lf2}). The best-fit Schechter functions of Equation\,\ref{eq:sch} are represented by dashed black lines, with the gray area representing $1\sigma$ uncertainties. The faint-end slope was determined at two low redshift bins, and the more accurate result, $\alpha=-0.26\pm 0.11$, was adopted at all the remaining redshifts, as explained in Section\,\ref{sec:lf4}.}
     \label{fig:phi}
\end{figure}

\begin{figure}
\centering
   \includegraphics[width=9cm]{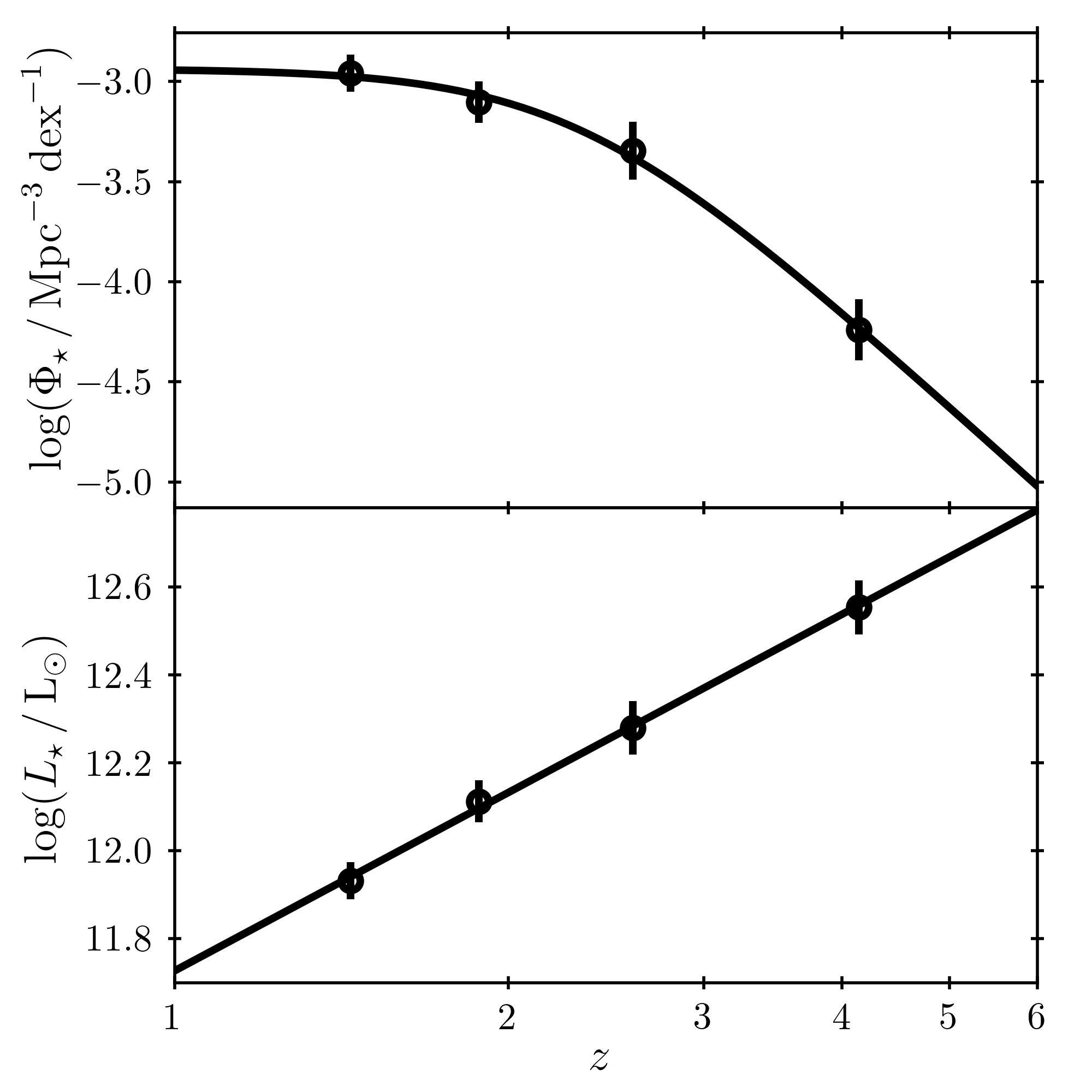}
     \caption{Redshift evolution of the Schechter function parameters found in this work. The values determined at each redshift (Table\,\ref{tab:pars}) are shown as black points, with the best-fit functions (Equations\,\ref{eq:parsf} and \ref{eq:parsfv}) depicted by black solid lines. (For details see Section\,\ref{sec:lf4}.)}
     \label{fig:pars}
\end{figure}

\begin{figure}
\centering
   \includegraphics[width=9cm]{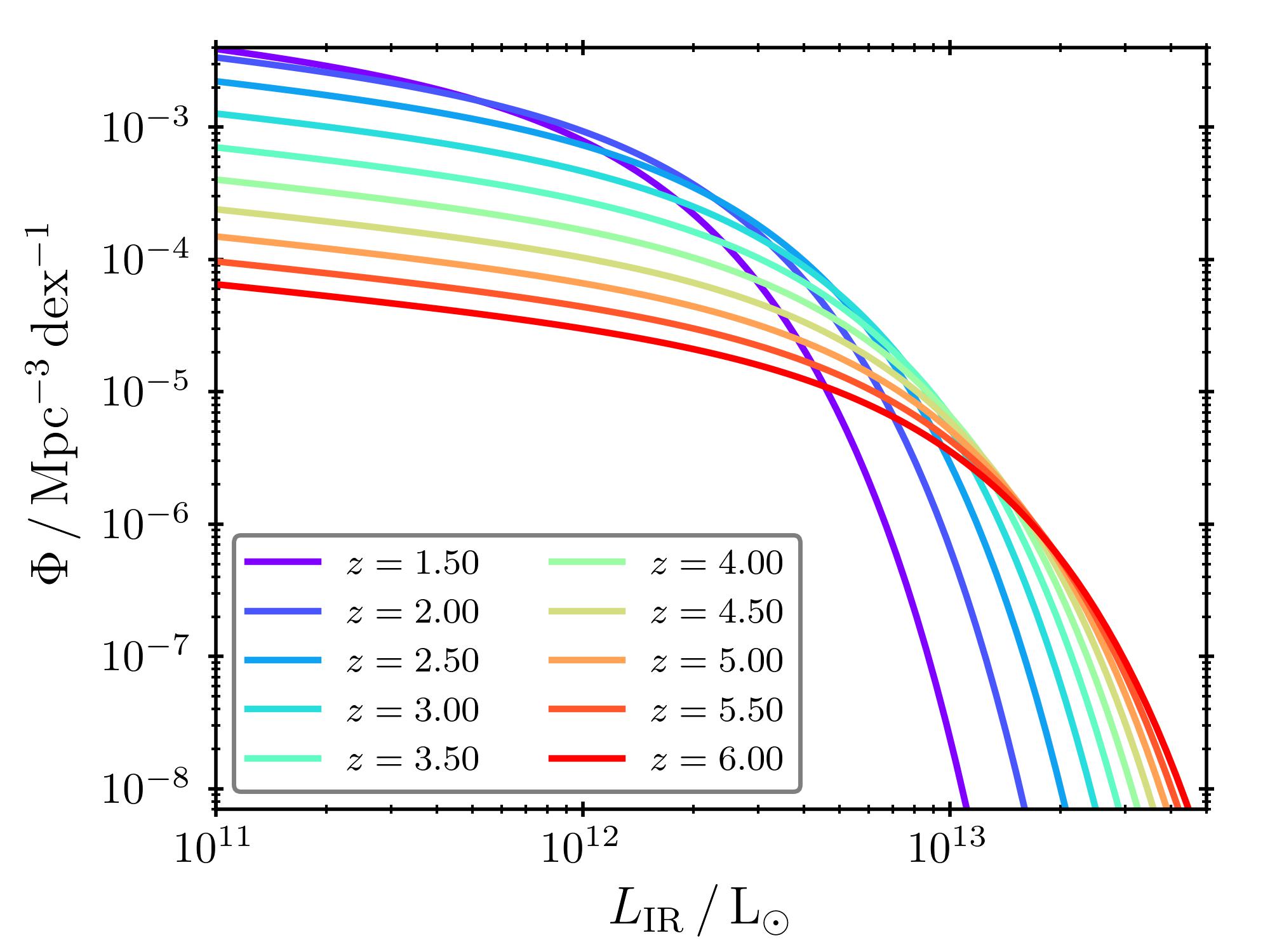}
     \caption{Redshift evolution of the functional form of the IR LF found in this work (Equations\,\ref{eq:sch}, \ref{eq:parsf} and \ref{eq:parsfv}).}
     \label{fig:lfall}
\end{figure}

Following previous works, we fit the IR LF with the Schechter function:

\begin{equation}\label{eq:sch}
\Phi(L,z)=\Phi_\star\left(\frac{L}{L_\star}\right)^\alpha{\rm exp}\left(\frac{-L}{L_\star}\right),
\end{equation}

\noindent where $\Phi_\star$ is the normalization parameter, $\alpha$ is the faint-end slope, and $L_\star$ is the characteristic luminosity that marks the border between the power law and the exponential fits (for comparison purposes, the modified Schechter fits are presented in Section\,\ref{sec:ap2}). Due to the relatively shallow depth of IR observations, the faint-end slope is usually determined only at redshifts $z\ll 1$ (e.g., \citealt{Gruppioni_2013}) and then fixed at higher redshifts, with the exception of \citet{Koprowski_2017}, where $\alpha$ was found using the ALMA survey of Hubble Ultra Deep Field \citep{Dunlop_2017} at $z\simeq 2$. Here, we fit the faint-end slope at the two lowest redshift bins ($1.2\leq z < 1.6$ and $1.6\leq z < 2.2$), finding $\alpha=-0.26\pm 0.11$ and $-0.23\pm 0.18$, respectively, and we adopted a more accurate $1.2\leq z < 1.6$ value of $-0.26$ at all redshift bins.

The faint- and bright-end data from Figure\,\ref{fig:phi} were fit at each redshift using Equation\,\ref{eq:sch} (solid black curves in Figure\,\ref{fig:phi}), with $\alpha$ set to $-0.26$ and using the nonlinear least squares analysis with the errors estimated from the corresponding covariance matrix. The resulting values of $\Phi_\star$ and $L_\star$ are summarized in Table\,\ref{tab:pars} and depicted as black points with error bars in Figure\,\ref{fig:pars}. The functional form for the redshift dependence for each parameter (black solid curves in Figure\,\ref{fig:pars}) was then found, where

\begin{equation}\label{eq:parsf}
\begin{split}
\Phi_\star/{\rm Mpc}^{-3}{\rm dex}^{-1} & = \Phi_{\star,0}\left[ \left( \frac{z}{z_0} \right)^{a_1} + \left( \frac{z}{z_0} \right)^{a_2} \right]^{-1} \\
L_\star/{\rm L_\odot} & = b_1 \times z^{b_2}
\end{split}
\end{equation}

\noindent and

\begin{equation}\label{eq:parsfv}
\begin{split}
\Phi_{\star,0} & = (1.22\pm 0.20) \times 10^{-3} \\
z_0 & = \phantom{(}2.24\pm 0.23 \\
a_1 & = \phantom{(}{\rm fixed} \\
a_2 & = \phantom{(}4.95\pm 0.73 \\
b_1 & = (5.26\pm 0.30) \times 10^{11} \\
b_2 & = \phantom{(}1.38 \pm 0.07,
\end{split}
\end{equation}

\begin{table}
\tiny
\caption{Schechter function (Equation \ref{eq:sch}) best-fit parameters.}\label{tab:pars}
\centering
\begin{tabular}{ccc}
\hline
<$z$> & ${\rm log}(\Phi_\star/{\rm Mpc}^{-3}{\rm dex}^{-1})$ & ${\rm log}(L_\star/{\rm L_\odot})$ \\
\hline
$1.44$ & $-2.98\pm 0.10$ & $11.93\pm 0.05$ \\
$1.88$ & $-3.13\pm 0.11$ & $12.13\pm 0.04$ \\
$2.59$ & $-3.33\pm 0.14$ & $12.28\pm 0.05$ \\
$4.14$ & $-4.23\pm 0.16$ & $12.55\pm 0.06$ \\
\hline
\end{tabular}
\tablefoot{The values were calculated at the center value of each redshift bin studied in this work (see Section \ref{sec:lf4} for details).}
\end{table}

\noindent with $a_1$ fixed at 0.0 following \citet{Casey_2018}.

\begin{figure*}
\centering
   \includegraphics[width=18cm]{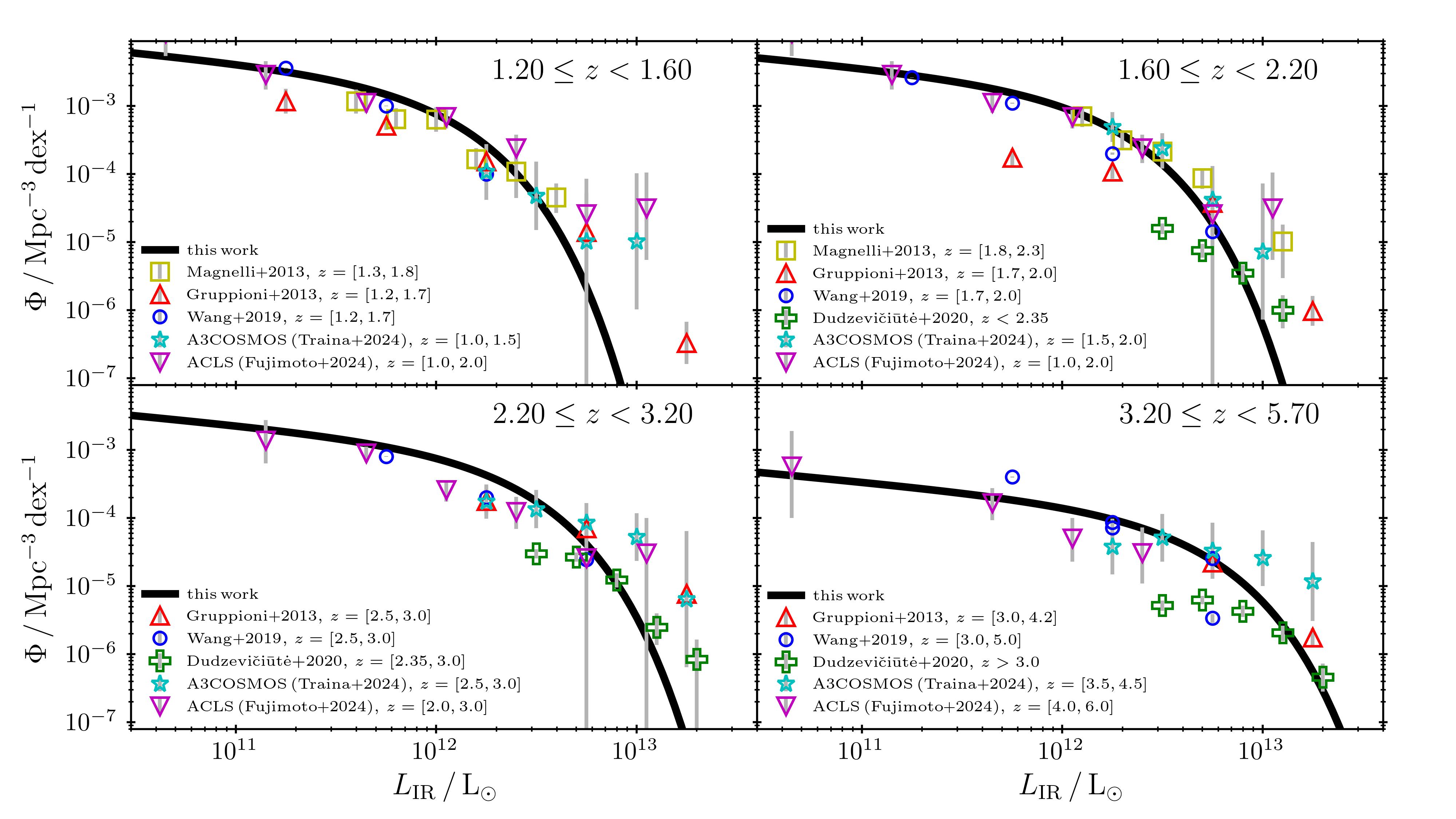}
     \caption{Functional form of the infrared luminosity function found in this work (black solid line). For comparison we show the individual data found in other literature studies (\citealt{Magnelli_2013, Gruppioni_2013, Wang_2019, Dudzeviciute_2020, Traina_2024}, and \citealt{Fujimoto_2024}). (For a detailed discussion, see Section\,\ref{sec:lf4}.)}
     \label{fig:phi1}
\end{figure*}

\begin{figure*}
\centering
   \includegraphics[width=18cm]{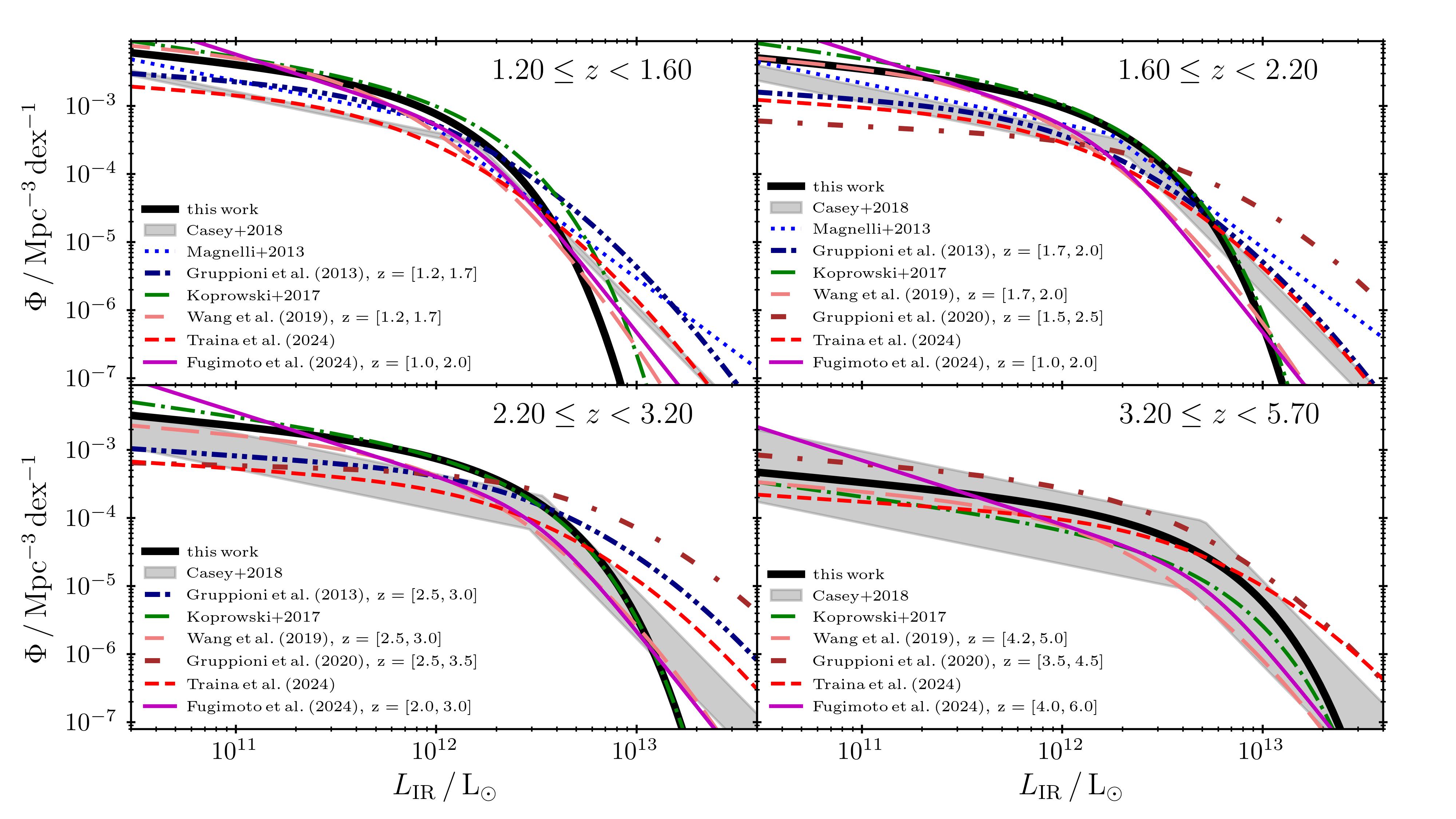}
     \caption{Functional form of the infrared luminosity function found in this work (black solid line). Other IR LFs found in the literature are also shown for comparison (\citealt{Magnelli_2013, Gruppioni_2013, Koprowski_2017, Wang_2019, Gruppioni_2020, Traina_2024}, and \citealt{Fujimoto_2024}). It can be seen that while at the faint end our results seem to be mostly consistent with other works (see Figure\,\ref{fig:phi1}), the assumed value of the faint-end slope causes the corresponding functions to differ significantly. The shaded region represents the regime bounded at the bottom by the dust-poor models and at the top by the dust-rich models of the early Universe postulated by \citet{Casey_2018}. (For a detailed discussion, see Section\,\ref{sec:lf4}.)}
     \label{fig:phi2}
\end{figure*}

In the top panel of Figure\,\ref{fig:pars} we present the redshift evolution of the characteristic number density, $\Phi_\star$, as a solid black line. The steep evolution found here, where ${\rm log}\,\Phi_\star \propto (1+z)^{-4.95}$ is close to the dust-poor model of \citet{Casey_2018}, who assumed ${\rm log}\,\Phi_\star \propto (1+z)^{-5.9}$. The redshift evolution of the characteristic luminosity, $L_\star$, is shown in the bottom panel of Figure\,\ref{fig:pars} with a solid black line. The increasing evolution of $L_\star$ with redshift is consistent with the downsizing scenario \citep{Thomas_2010} in which the more luminous (and more massive) galaxies formed earlier than their fainter counterparts (this is also apparent in Figure\,\ref{fig:lfall}).

In Figures\,\ref{fig:phi1} and \ref{fig:phi2}, we present a comparison of our functional form of the IR LF with other data and their corresponding best-fit functions from the literature \citep{Magnelli_2013, Gruppioni_2013, Koprowski_2017, Casey_2018, Dudzeviciute_2020, Traina_2024, Fujimoto_2024}. Figure\,\ref{fig:phi1} shows that our results agree (within the errors) with the majority of the studies shown here, with notable exceptions being the {\it Herschel}-selected samples from \citet{Gruppioni_2013} and the inhomogeneous ALMA COSMOS data presented by \citet{Traina_2024}. Previous works include discussion of the discrepancies between submillimeter-selected and {\it Herschel}-selected samples (e.g., \citealt{Koprowski_2017}), where the low-resolution {\it Herschel} data from \citet{Gruppioni_2013} is affected by blending issues, which likely leads to an overestimation of the infrared luminosity values, where for the brightest {\it Herschel} sources the 250 ${\rm \mu m}$ flux may be overestimated by around 150\% \citep{Scudder_2016} and is also prone to active galactic nucleus contamination. The ALMA data used in \citet{Traina_2024} consists of the publicly available COSMOS data and is therefore inhomogeneous in terms of observing wavelength and depth. In addition, ALMA observations tend to target the most luminous IR sources, and hence, due to clustering, the IR LFs constructed from such data are likely biased toward larger number densities. \citet{Traina_2024} aimed at fixing the clustering bias of their sample by excluding central sources in all of their ALMA pointings. However, the ALMA data, by selection, target relatively dense regions in the field, likely boosting the resulting values of the IR LF. The works of \citet{Magnelli_2013} and \citet{Fujimoto_2024} largely agree with our results (within errors), with the exception of the brightest $L_{\rm IR}$ bin at $z\sim 2$. In the case of \citet{Magnelli_2013}, the sample was selected at the {\it Herschel} PACS wavelength, with the {\it Herschel} fluxes affected by blending and the PACS selection bands likely contaminated by the active galactic nuclei. In the case of \citet{Fujimoto_2024}, the sample studied was based on the 180 dust continuum sources identified in 33 massive cluster fields and hence can also likely be biased toward larger number densities at the bright end. On the other hand, the 850 ${\rm \mu m}$ selection of this work will potentially miss low-dust mass, high-dust temperature sources. Although we tried to account for this effect in our completeness calculations, some sources with a low dust mass and relatively high dust temperature may still be missed in our sample. 

In Figure\,\ref{fig:phi1} we also plot the results of \citet{Dudzeviciute_2020} with green crosses, where the IR LF values were determined at IR luminosities above the limit corresponding to the $S_{870}$ flux of 3.6\,mJy and the median temperature of the AS2UDS sample of $\sim 30$\,K without the requirement of at least one SPIRE detection. Since we assumed a more conservative dust temperature value of 37.3\,K, we effectively limited our analysis to the most luminous IR bins (Section\,\ref{sec:lf2}). As the submillimeter flux roughly traces the dust mass of the galaxies ($M_{\rm d}$; e.g., \citealt{Dudzeviciute_2020}), our lower-limit IR luminosity bins will statistically miss a small fraction of sources with ${\rm log}(M_{\rm d}/{\rm M_\odot})\sim 9.0$ ($S_{870}\sim 4$\,mJy) and $T_{\rm d}>37.3$\,K (increasing to higher dust temperatures at $z\gtrsim 4$ due to SPIRE detection requirement). Assuming a lower dust temperature produces less luminous IR lower limits, which become increasingly less complete. The small differences at the highest $L_{\rm IR}$ bins between both works can mainly be attributed to the different redshift bins adopted but also to the fact that the sample used in \citet{Dudzeviciute_2020} included sources without any {\it Herschel} SPIRE detections and did not adopt the completeness corrections of Equation\,\ref{eq:lfbe} from the original SCUBA-2 sample of \citet{Geach_2017}.

The IR LF functional forms corresponding to the data from Figure\,\ref{fig:phi1} presented in Figure\,\ref{fig:phi2} show the impact of the assumed faint-end slope on the resulting form of the IR LF. Even though, as can be seen in Figure\,\ref{fig:phi1}, the literature data are mostly consistent with our findings at the faint end, the assumed s$\alpha$ impacts the resulting functional form significantly, which in turn slightly affects the corresponding values of the star formation rate density, $\rho_{\rm SFR}$, which we discuss in the next subsection.

\subsection{Cosmic star formation history} \label{sec:sfrd}

\begin{figure*}[h!]
\centering
   \includegraphics[width=18cm]{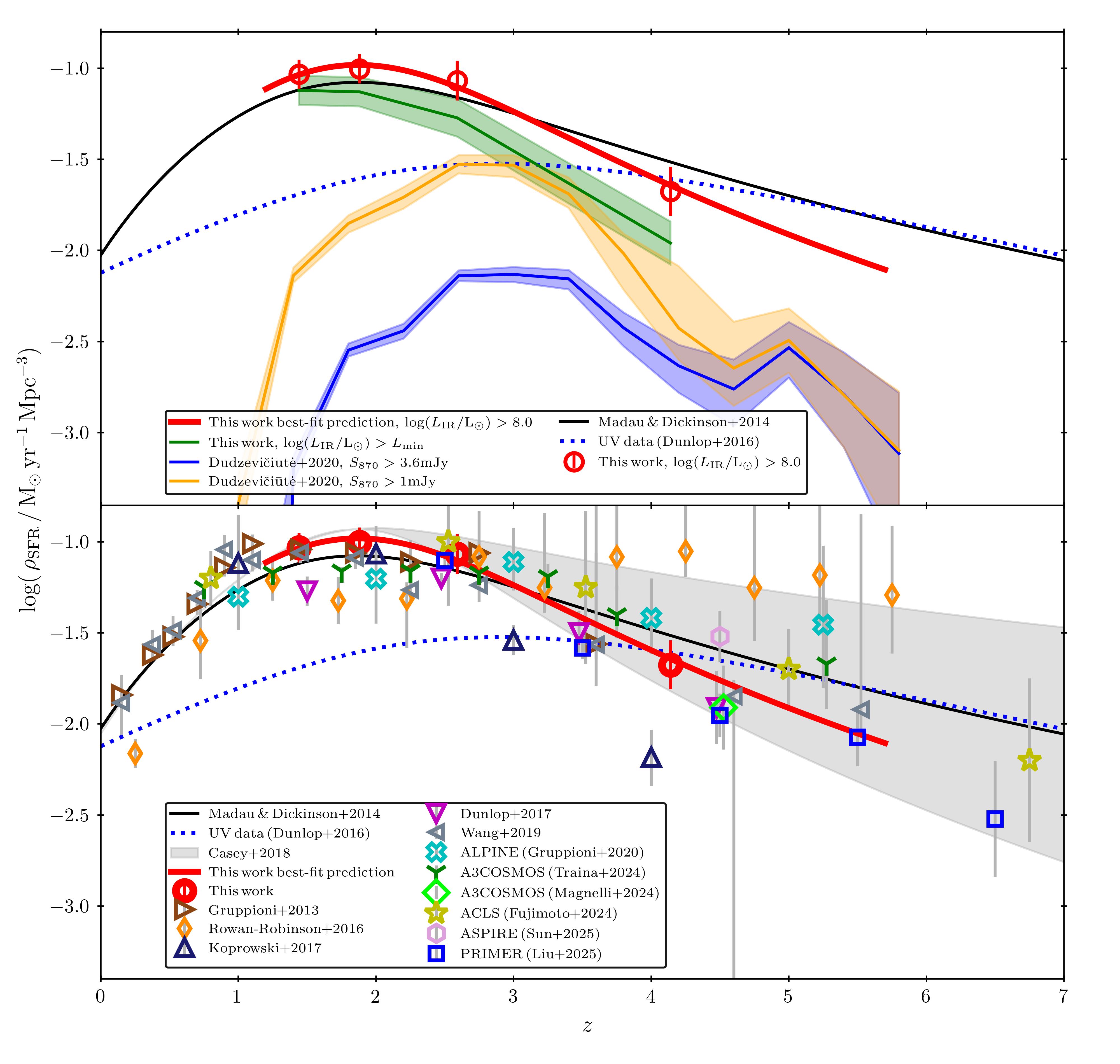}
     \caption{Star formation rate density for four redshift bins studied in this work (Table\,\ref{tab:sfrd}), depicted with red circles with error bars, found by integrating the corresponding IR LFs between $8.0<L_{\rm IR}/{\rm L_\odot}<14.0$. The solid red line represents the redshift evolution of the star formation rate density derived from the best fits to the Schechter function parameters of Figure\,\ref{fig:pars} between redshifts considered in this work ($1.2<z<5.7$). The solid black line represents the IR $\rho_{\rm SFR}$ evolution found in \citet{Madau_2014}, while the dotted blue line shows the evolution of the UV $\rho_{\rm SFR}$ of \citet{Dunlop_2016}. \textbf{Top:} Comparison of our results with those presented in \citet{Dudzeviciute_2020}, where the blue and orange lines depict the contribution to $\rho_{\rm SFR}$ from 870\,${\rm \mu m}$ sources with fluxes above 3.6\,mJy and 1.0\,mJy (extrapolated using ALMA number counts of \citealt{Hatsukade_2018}), respectively. The shaded areas represent 1$\sigma$ errors. The green line shows the evolution down to the IR luminosity detection limits of our IR faint sample presented in Section\,\ref{sec:lf1}. The differences in the derived values of the $\rho_{\rm SFR}$ at a specific redshift across the samples arise from the different $L_{\rm IR}$ limits adopted (see Section\,\ref{sec:sfrd}). At $z\gtrsim 4$, these differences decrease, which can be attributed to the growing contribution of bright submillimeter galaxies to the $\rho_{\rm SFR}$. \textbf{Bottom:} Comparison of $\rho_{\rm SFR}$ found in this work with the most recent results from the literature plotted with color symbols \citep{Gruppioni_2013, Rowan_2016, Koprowski_2017, Dunlop_2017, Wang_2019, Gruppioni_2020, Traina_2024, Magnelli_2024, Fujimoto_2024, Sun_2025, Liu_2025}. The shaded region shows the results of \citet{Casey_2018} bounded at the bottom by the dust-poor models and at the top by the dust-rich models of the early Universe. It can be seen that our results are in a good agreement with the results of \citet{Dunlop_2017, Wang_2019, Magnelli_2024}, and \citet{Liu_2025}. Also, some inconsistencies with other works are apparent, the discussion of which is presented in Section\,\ref{sec:sfrd}. Considering the shaded region of \citet{Casey_2018}, it can be seen that the results of this work point toward the dust-poor early Universe scenario, where the IR sources are expected to dominate the total $\rho_{\rm SFR}$ out to $z\sim 4$.}
     \label{fig:sfrd}
\end{figure*}

In order to find the comoving volume density of the IR luminosity, $\rho_{\rm IR}$, at redshift $z_i$, we followed

\begin{equation}\label{eq:rhoir}
   \rho_{\rm IR}(z_i)=\int_{{\rm log}(L_{\rm min})}^{{\rm log}(L_{\rm max})}\Phi(L,z_i)\times L\,{\rm dlog}(L),
\end{equation}

\noindent where $\Phi$ is the Schechter function of Equation\,\ref{eq:sch}, with the best-fit parameters at $z_i$ listed in Table\,\ref{tab:pars}. Following previous works (e.g., \citealt{Koprowski_2017, Gruppioni_2013, Traina_2024}), we integrate Equation\,\ref{eq:rhoir} between $8.0<{\rm log}(L_{\rm IR}/{\rm L_\odot})<14.0$. To find the star formation rate density, we then followed \citet{Kennicutt_1998}:

\begin{equation}\label{eq:rhosfr}
    \rho_{\rm SFR} = \rho_{\rm IR} \times \mathcal{K}_{\rm IR} \times 0.63,
\end{equation}

\noindent with $\mathcal{K}_{\rm IR}=1.73\times 10^{-10}\,{\rm M_\odot\, year^{-1}\,L_\odot^{-1}}$ and an additional multiplicative factor of 0.63 to convert from a Salpeter to a Chabrier IMF. The resulting values are summarized in Table\,\ref{tab:sfrd} and depicted as red circles with error bars in Figure\,\ref{fig:sfrd}. The red solid line depicts the redshift evolution of the star formation rate density derived from the best-fit functions of Figure\,\ref{fig:pars} between the redshifts considered in this work ($1.2<z<5.7$). In order to estimate the errors on $\rho_{\rm SFR}$ at each redshift listed in Table\,\ref{tab:pars}, we performed Monte Carlo simulations, where in each of the 1000 realizations, the IR luminosities and their corresponding IR LF values listed in Tables\,\ref{tab:fe} and \ref{tab:be} were randomly sampled from a Gaussian distribution with the mean and the variance equal to the calculated values and their errors, respectively. In each of the realizations, the best-fit Schechter function was found (Equation\,\ref{eq:sch}), and the corresponding $\rho_{\rm SFR}$ was determined following Equations\,\ref{eq:rhoir} and \ref{eq:rhosfr}. The final errors on $\rho_{\rm SFR}$ at each redshift were then taken to be equal to the standard deviations of all the individual values.

\begin{table}
\tiny
\caption{Star formation rate density values at the redshift bins studied in this work.}\label{tab:sfrd}
\centering
\begin{tabular}{cc}
\hline
$z_{\rm center}$ & ${\rm log}(\rho_{\rm SFR}/{\rm M_\odot\,yr^{-1}\,Mpc^{−3}})$ \\
\hline
1.44 & $-1.04\pm 0.07$ \\
1.88 & $-0.98\pm 0.08$ \\
2.59 & $-1.07\pm 0.09$ \\
4.14 & $-1.71\pm 0.13$ \\
\hline
\end{tabular}
\end{table}

In the top panel of Figure\,\ref{fig:sfrd}, we compare our results with those presented in \citet{Dudzeviciute_2020}. The blue line (with 1$\sigma$ errors represented by the shaded area) depicts the results presented in \citet{Dudzeviciute_2020} down to the $S_{870}$ flux limit of 3.6\,mJy, while the orange line shows the derived values extrapolated down to the $S_{870}$ flux limit of 1.0\,mJy, where the ALMA number counts of \citet{Hatsukade_2018} were used. The green line gives the evolution of $\rho_{\rm SFR}$ found in this work down to the IR luminosity detection limits of our IR faint sample presented in Section\,\ref{sec:lf1} (Table\,\ref{tab:fe}), and the red circles with 1$\sigma$ error bars represent the integrated values down to ${\rm log}(L_{\rm IR}/L_\odot)=8.0$, with the red solid line produced from the evolution of the best-fit Schechter function parameters, presented in Section\,\ref{sec:lf4}. The increasing values of $\rho_{\rm SFR}$ at $z\lesssim 3$ for each of the samples are simply a consequence of different lower $L_{\rm IR}$ limits adopted, where $S_{870}$ fluxes from \citet{Dudzeviciute_2020} of 3.6 and 1.0\,mJy correspond to a ${\rm log}(L_{\rm IR}/L_\odot)$ of $\sim 12.4$ and $\sim 11.7$, respectively. For the sample of this work represented by the green line in the figure, the IR LF lower limit is set to a ${\rm log}(L_{\rm IR}/L_\odot)$ of $\sim 10.9$ at $z\sim 1.4$, rising to $\sim 12.0$ at $z>4$ (Table\,\ref{tab:fe}), while the red circles were produced by integrating the best-fit Schechter functions down to ${\rm log}(L_{\rm IR}/L_\odot)=8.0$. Interestingly, the differences in the derived $\rho_{\rm SFR}$ values for each sample decrease toward higher redshifts, which indicates the growing contribution of the most IR luminous sources to the total star formation rate density toward earlier epochs (which is also apparent in Figure\,\ref{fig:lfall}).

In the bottom panel of Figure\,\ref{fig:sfrd}, we compare our findings with a range of recent results from the literature, where whenever possible, the conversion from $L_{\rm IR}$ to SFR was recalibrated using Equation\,\ref{eq:rhosfr}. It can be seen that the numbers found in this work are in a good agreement with those determined in \citet{Dunlop_2017}, \citet{Wang_2019}, \citet{Magnelli_2024}, and \citet{Liu_2025}. In the case of \citet{Dunlop_2017} and \citet{Liu_2025}, the cosmic star formation rate density was derived from individual ALMA detections. In order to account for sources that went undetected by ALMA, \citet{Dunlop_2017} performed stacking down to the stellar mass limit of $\sim 10^{9.3}\,{\rm M_\odot}$, while \citet{Liu_2025} included the contribution from the ALMA-undetected galaxies down to $M_\ast>10^{10}\,{\rm M_\odot}$.  \citet{Magnelli_2024} estimated the $\rho_{\rm SFR}$ at $z\sim 4.5$ by $uv$-plane stacking COSMOS sources in the available ALMA data down to $M_\ast=10^{9.5}\,{\rm M_\odot}$. These lower limits are consistent with the recent findings in which no dust was detected in high-redshift galaxies with stellar masses below $M_\ast \sim 10^{9.5}\,{\rm M_\odot}$ (e.g., \citealt{Pannella_2015, McLure_2018, Koprowski_2018}). 

It can also be seen in Figure\,\ref{fig:sfrd} that our results are somewhat lower than those of \citet{Traina_2024,Fujimoto_2024,Sun_2025} and significantly lower than the results found in \citet{Rowan_2016} and \citet{Gruppioni_2020} at $z\gtrsim 3$. At high redshifts, the selected 350 ${\rm \mu m}$ and 500 ${\rm \mu m}$ {\it Herschel} sample of \citet{Rowan_2016} consists only of the sources with the most extreme S/N (${\rm S/N}>5$), and the sources have IR luminosities considerably higher than the knee of the corresponding LF. \citet{Gruppioni_2020}, as can be seen in Figure\,\ref{fig:phi2}, predicts bright-end number densities that are significantly larger than what was found in this work. In fact, their IR LF suggests approximately ten times more sources at the bright end than what is predicted by the somewhat extreme case of the dust-rich early Universe scenario of \citet{Casey_2018}.

The shaded region in Figure\,\ref{fig:sfrd} depicts the $\rho_{\rm SFR}$ theoretical regime bounded from the bottom by the dust-poor scenario and from the top by the dust-rich scenario of the early Universe, as calibrated by \citet{Casey_2018}. In the dust-poor model, the UV-bright sources (dotted blue line in Figure\,\ref{fig:sfrd}; \citealt{Dunlop_2016}) dominate star formation at $z\gtrsim 3.5$, where the dust-formation timescale, driven primarily by asymptotic giant branch (AGB) stars and supernovae, is expected to be longer than the time it takes to form the first UV-bright galaxies. The dust-rich model, on the other hand, predicts the cosmic star formation to be dominated by the dusty star forming galaxies between $1.5<z<6.5$. As noted by \citet{Casey_2018}, the dust-rich model is somewhat extreme, as very few DSFGs have been found at $z\gtrsim 5$, mainly due to observational limitations (see \citealt{Casey_2014} for details). Indeed, from Figure\,\ref{fig:sfrd} it can be seen that the IR LF found in this work predicts number densities of DSFGs that are similar to the dust-poor model and consistent with the low-detection rates at high redshifts, placing our results in a dust-poor early Universe regime, where the IR sources seem to dominate the star formation rate density at least out to $z\sim 4$.

\section{Summary} \label{sec:sum}

We have probed the low-luminosity regime of the IR luminosity function out to $z=5.7$ by stacking the optical/near-IR catalogs of \citet{McLeod_2021} in the FIR {\it Herschel} and JCMT maps, using stellar mass as a proxy, where the $L_{\rm IR}$-$M_\ast$ relationship of \citet{Koprowski_2024} was used. Together with the ALMA follow-up data of the S2CLS UKIDSS UDS sources (AS2UDS; \citealt{Stach_2019, Dudzeviciute_2020}), we have established the evolution of the functional form of IR LF between redshifts 1.2 and 5.7. We fit the data at four redshift bins (Table\,\ref{tab:pars}) with the Schechter function of Equation\,\ref{eq:sch}, and we established a value of the faint-end slope; the redshift evolution of the characteristic luminosity, $L_\ast$; and the characteristic number density, $\Phi_\ast$. By integrating the best-fit functions, we have determined the values of the comoving volume density of star formation, $\rho_{\rm SFR}$. The main results of this work can be summarized as follows:

\begin{enumerate}[label=(\roman*),wide]

\item{The faint end of the IR luminosity function values of $\alpha=-0.26\pm 0.11$ and $-0.23\pm 0.18$ were determined at two low-redshift bins of $1.2\leq z < 1.6$ and $1.6\leq z < 2.2$, respectively, and the more accurate $1.2\leq z < 1.6$ value of -0.26 was adopted at all remaining redshift bins. The characteristic number density at $z\gtrsim 2$ is a steeply declining function of redshift, where $\Phi_\ast\propto z^{-4.95}$, and it is very similar to the predicted evolution for the dust-poor model of the early Universe investigated in \citet{Casey_2018}. The redshift evolution of the characteristic luminosity was found to be a power-law function, where $L_\star\propto z^{1.38}$, supporting the downsizing scenario \citep{Thomas_2010}, where most luminous galaxies form before their less luminous counterparts.}

\item{When comparing the functional form of the IR LF with the results from the recent literature, we found our faint-end slope to be in agreement with most of other calibrations. At the bright end, we predict relatively small number densities of sources, which aligns with recent studies struggling to detect high-redshift galaxies in the FIR \citep{Dunlop_2017,Aravena_2020}. A small number of previous studies found significantly more high-$L_{\rm IR}$ sources, which we attribute to the low resolution of the FIR observations and/or the methodology adopted when dealing with inhomogeneous FIR samples.}

\item{Finally, the redshift evolution of the star formation rate density, $\rho_{\rm SFR}$, was established by integrating the corresponding IR LFs at the four redshift bins investigated in this work. The resulting $\rho_{\rm SFR}$ peaks at $z\simeq 2$ and later declines with redshift from $z\sim 2$ out to $z\sim 6$, with the high-redshift values significantly lower than those presented in \citet{Madau_2014}. Our data are in very good agreement with the recent works of \citet{Dunlop_2017}, \citet{Wang_2019}, \citet{Magnelli_2024}, and \citet{Liu_2025}, but they are somewhat below the numbers derived in \citet{Traina_2024,Fujimoto_2024} and \citet{Sun_2025}. The most striking differences are between our results and those established by \citet{Rowan_2016} and \citet{Gruppioni_2020}, who significantly larger number densities have been found at the bright end. Considering the two extreme cases of a dust-rich and a dust-poor early Universe presented in \citet{Casey_2018}, the $\rho_{\rm SFR}$ redshift evolution found in this work places our findings in the dust-poor scenario, where we predict that the IR sources will dominate the total density of star formation in the Universe out to $z\sim 4$.}

\end{enumerate}

\begin{acknowledgements}

This research was funded in whole or in part by the National Science Center, Poland (grants no. 2020/39/D/ST9/03078, 2023/51/B/ST9/01479, 2023/49/B/ST9/00066 and 2024/53/N/ST9/00350). For the purpose of Open Access, the author has applied a CC-BY public copyright license to any Author Accepted Manuscript (AAM) version arising from this submission. JSD and DJM acknowledge the support of the Royal Society through a Research Professorship awarded to JSD. KL acknowledges the support of the National Science Centre, Poland, through the PRELUDIUM grant UMO-2023/49/N/ST9/00746. Supported by the Foundation for Polish Science (FNP).

\end{acknowledgements}

\bibliographystyle{aa}    
\bibliography{papers}

\newpage

\begin{appendix}

\section{Modified-Schechter fits}\label{sec:ap2}

\begin{figure}
\centering
   \includegraphics[width=9cm]{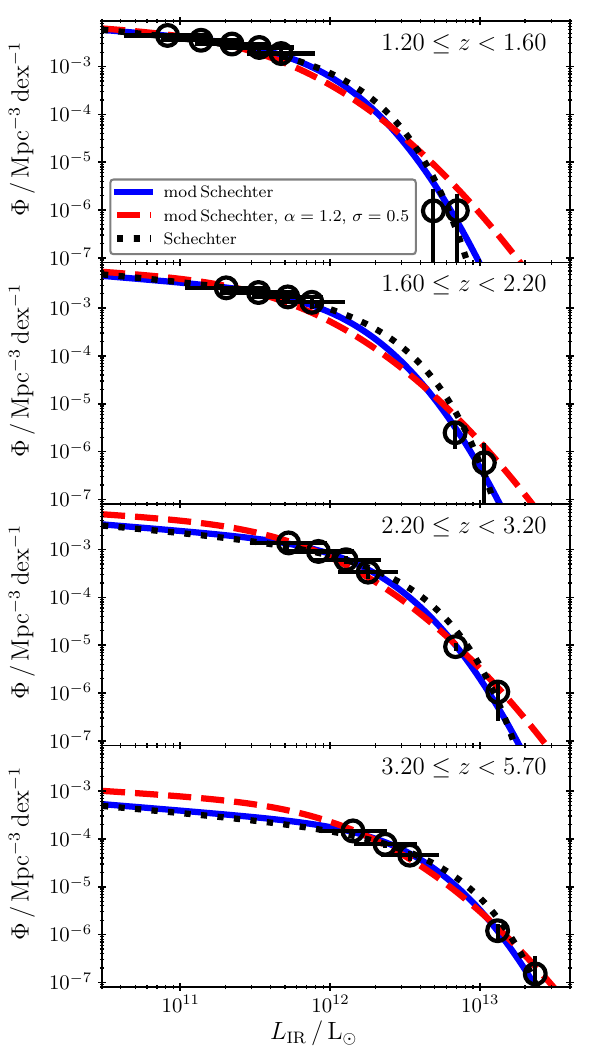}
     \caption{Infrared LF binned data of Figure\,\ref{fig:phi} (black circles with error bars) with the best-fit functional forms overlayed. The modified Schechter fits with $\alpha$ and $\sigma$ parameters (Equation\,\ref{eq:modsch}) allowed to vary freely are depicted with blue solid line. Fixing $\alpha$ and $\sigma$ at the recent literature values of -0.2 and 0.5, respectively \citep{Gruppioni_2013,Wang_2019,Gruppioni_2020,Traina_2024}, produces the modified Schechter function presented with the red dashed line, while the best-fit classical Schechter curve (Equation\,\ref{eq:sch}) derived in this work is shown with the black dotted line.}
     \label{fig:lfmsch}
\end{figure}

\begin{figure}
\centering
   \includegraphics[width=9cm]{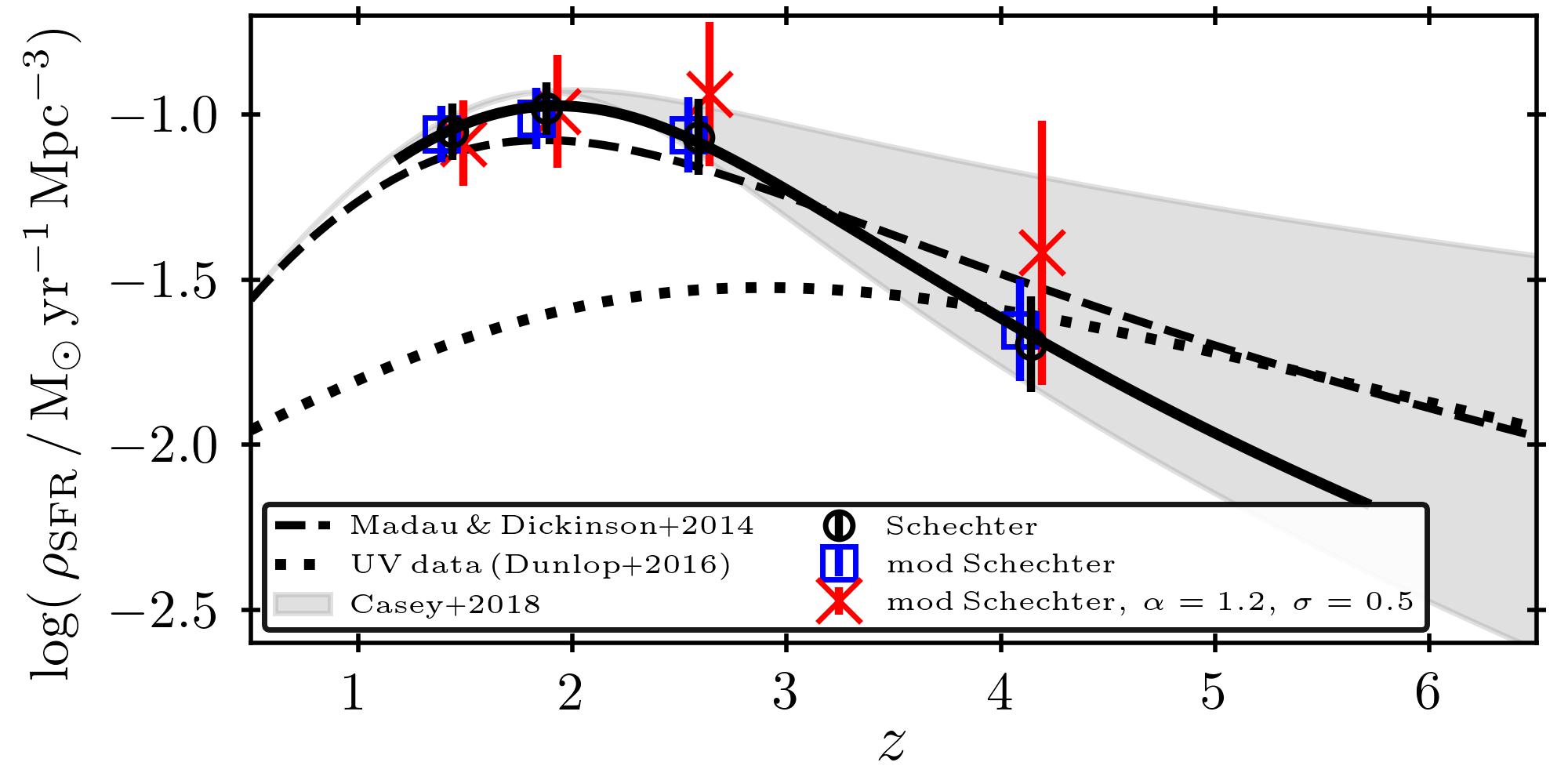}
     \caption{Star formation rate density found by integrating the best-fit curves of Figure\,\ref{fig:lfmsch} between $8.0<L_{\rm IR}/{\rm L_\odot}<14.0$. Similarly to Figure\,\ref{fig:sfrd}, the black solid line represents the IR $\rho_{\rm SFR}$ evolution found in \citet{Madau_2014}, while the blue dotted line shown the evolution of the UV $\rho_{\rm SFR}$ of \citet{Dunlop_2016}. The shaded region shows the results of \citet{Casey_2018}, bounded at the bottom by the dust-poor models and at the top by the dust-rich models of the early Universe. The black circles with the solid curve represent this work's results found in Section\,\ref{sec:sfrd}. Values of $\rho_{\rm SFR}$ corresponding to the modified Schechter fits (Eqaution\,\ref{eq:modsch}) with the best-fit $\alpha$ and $\sigma$ parameters of $-0.26$ and 0.31, respectively, are shown with blue squares. Data found from modified Schechter fits with $\alpha$ and $\sigma$ fixed at recent literature $z\ll 1$ values of $-0.2$ and 0.5, respectively, are depicted with red crosses. It can be seen that, while using the modified Schechter function with all the parameters allowed to vary freely produces results almost identical to the classical Schechter results of this work, setting $\alpha$ and $\sigma$ at the $z\ll 1$ best-fit values if \citet{Gruppioni_2013}, produces significantly larger numbers at highest redshift bins.}
     \label{fig:sfrdmsch}
\end{figure}

Since a number of recent works have adopted a modified version of the Schechter function (e.g., \citealt{Gruppioni_2013,Wang_2019,Gruppioni_2020,Traina_2024}), in this appendix we present an alternative IR LF functional fits and the corresponding star formation rate density calculations following the original work of \citet{Saunders_1990}, where

\begin{equation}\label{eq:modsch}
    \Phi(L) = \Phi_\ast\left(\frac{L}{L_\ast}\right)^{\alpha}{\rm exp}\left[-\frac{1}{2\sigma^2}{\rm log}^2_{10}\left(1+\frac{L}{L_\ast}\right)\right].
\end{equation}

The free parameters $\Phi_\ast$ and $L_\ast$ mark the so-called knee of the luminosity function, with $L\ll L_\ast$ behaving as a power-law and $L\gg L_\ast$ following a Gaussian curve. This function has proven to be useful, since it allows the control of the shape of the bright-end portion of the curve through the $\sigma$ parameter. While the classical Schechter function adopted in this work does not require any modifications at the bright end, it is interesting to investigate whether the modified version produces different results in terms of $\rho_{\rm SFR}$. In most of the high-redshift studies (e.g., \citealt{Gruppioni_2013,Wang_2019,Gruppioni_2020,Traina_2024}), both $\alpha$ and $\sigma$ are fixed at $z\ll 1$ best-fit values of $-0.2$ and 0.5, respectively. In Figure\,\ref{fig:lfmsch} we show the binned data of Figure\,\ref{fig:phi} with black circles with the best-fit classical Schechter function found in this work depicted with black dotted lines. The modified Schechter fits, were $\alpha$ and $\sigma$ are fixed at $-0.2$ and 0.5, respectively, are presented with red dashed lines, while the similar fits with $\alpha$ and $\sigma$ set as free parameters are shown with blue solid lines. It can be seen that fixing $\alpha$ and $\sigma$ at $z\ll1$ values produces very poor fits at two lowest redshift bins, where the faint-end portion of the curve is best traced by the data. Allowing those parameters to vary (blue solid lines in Figure\,\ref{fig:lfmsch}) given significantly better fits, with the best-fit values of $\alpha=-0.26$ and $\sigma=0.31$. 

The corresponding star formation rate densities are shown in Figure\,\ref{fig:sfrdmsch}, with the black circles and the solid black line representing the results of this work presented in Section\,\ref{sec:sfrd}. The blue squares, showing the $\rho_{\rm SFR}$ values derived from the modified Schechter fits with $\alpha$ and $\sigma$ set as free parameters (blue solid lines in Figure\,\ref{fig:lfmsch}), are in an excellent agreement with the classical Schechter fits adopted in this work, while the data corresponding to the $z\ll 1$ low-resolution {\it Herschel} modified Schechter parameters found in \citet{Gruppioni_2013}, give significantly higher values of $\rho_{\rm SFR}$ at high-redshift bins.

\begin{sidewaystable*}
\section{Additional table}\label{sec:ap1}
\tiny
\caption{$L_{\rm IR}/{\rm L_\odot}$ stacking results.}\label{tab:lir}
\begin{tabular}{ccccccccc}
\hline\hline
& $0.45\leq z<0.60$ & $0.60\leq z<0.75$ & $0.75\leq z<1.00$ & $1.00\leq z<1.25$ & $1.25\leq z<1.60$ & $1.60\leq z<2.20$ & $2.20\leq z<3.20$ & $3.20\leq z<5.70$ \\ 
& log($L_{\rm IR}/{\rm L_\odot}$) & log($L_{\rm IR}/{\rm L_\odot}$) & log($L_{\rm IR}/{\rm L_\odot}$) & log($L_{\rm IR}/{\rm L_\odot}$) & log($L_{\rm IR}/{\rm L_\odot}$) & log($L_{\rm IR}/{\rm L_\odot}$) & log($L_{\rm IR}/{\rm L_\odot}$) & log($L_{\rm IR}/{\rm L_\odot}$) \\
\hline
$\phantom{0}9.25\leq \mathcal{M} < \phantom{0}9.50$ & $10.04\pm 0.07$ & $10.29\pm 0.07$ & $10.38\pm 0.08$ & $10.56\pm 0.07$ & & & & \\ 
$\phantom{0}9.50\leq \mathcal{M} < \phantom{0}9.75$ & $10.29\pm 0.06$ & $10.55\pm 0.06$ & $10.63\pm 0.06$ & $10.80\pm 0.06$ & $10.93\pm 0.07$ & & & \\ 
$\phantom{0}9.75\leq \mathcal{M} < 10.00$ & $10.56\pm 0.06$ & $10.77\pm 0.06$ & $10.89\pm 0.06$ & $10.99\pm 0.06$ & $11.20\pm 0.06$ & $11.32\pm 0.06$ & & \\ 
$10.00\leq \mathcal{M} < 10.25$ & $10.69\pm 0.06$ & $10.93\pm 0.06$ & $11.09\pm 0.05$ & $11.19\pm 0.05$ & $11.37\pm 0.06$ & $11.52\pm 0.06$ & $11.70\pm 0.06$ & \\ 
$10.25\leq \mathcal{M} < 10.50$ & $10.84\pm 0.05$ & $11.01\pm 0.06$ & $11.21\pm 0.05$ & $11.37\pm 0.05$ & $11.54\pm 0.05$ & $11.72\pm 0.05$ & $11.82\pm 0.06$ & $12.26\pm 0.08$ \\ 
$10.50\leq \mathcal{M} < 10.75$ & $10.84\pm 0.06$ & $11.14\pm 0.06$ & $11.33\pm 0.05$ & $11.48\pm 0.05$ & $11.67\pm 0.05$ & $11.80\pm 0.05$ & $12.00\pm 0.06$ & $12.36\pm 0.10$ \\ 
$10.75\leq \mathcal{M} < 11.00$ & $11.09\pm 0.06$ & $11.19\pm 0.06$ & $11.34\pm 0.06$ & $11.59\pm 0.06$ & $11.73\pm 0.06$ & $11.98\pm 0.06$ & $12.18\pm 0.06$ & $12.48\pm 0.09$ \\ 
$11.00\leq \mathcal{M} < 11.25$ & $11.04\pm 0.12$ & $11.25\pm 0.09$ & $11.51\pm 0.08$ & $11.61\pm 0.07$ & $11.88\pm 0.07$ & $12.06\pm 0.06$ & $12.38\pm 0.06$ & $12.72\pm 0.08$ \\ 
$11.25\leq \mathcal{M} < 11.50$ & \phantom{0}-- & $11.13\pm 0.38$ & $11.45\pm 0.15$ & $11.82\pm 0.10$ & \phantom{0}-- & $12.37\pm 0.08$ & $12.51\pm 0.10$ & $12.84\pm 0.09$ \\ 
\hline
 & N of sources & N of sources & N of sources & N of sources & N of sources & N of sources & N of sources & N of sources \\
\hline
$\phantom{0}9.25\leq \mathcal{M} < \phantom{0}9.50$  & $\phantom{0}1287\pm 36\phantom{00}$ & $\phantom{0}1868\pm 40\phantom{00}$ & $\phantom{0}3983\pm 56\phantom{00}$ & $\phantom{0}4647\pm 69\phantom{00}$ & & & & \\ 
$\phantom{0}9.50\leq \mathcal{M} < \phantom{0}9.75$  & $\phantom{0}1033\pm 32\phantom{00}$ & $\phantom{0}1415\pm 37\phantom{00}$ & $\phantom{0}2967\pm 51\phantom{00}$ & $\phantom{0}3378\pm 51\phantom{00}$ & $\phantom{0}4347\pm 63\phantom{00}$ & & & \\ 
$\phantom{0}9.75\leq \mathcal{M} < 10.00$  & $\phantom{00}769\pm 26\phantom{00}$ & $\phantom{00}983\pm 27\phantom{00}$ & $\phantom{0}2172\pm 46\phantom{00}$ & $\phantom{0}2503\pm 49\phantom{00}$ & $\phantom{0}3237\pm 56\phantom{00}$ & $\phantom{0}4259\pm 61\phantom{00}$ & & \\ 
$10.00\leq \mathcal{M} < 10.25$  & $\phantom{00}507\pm 22\phantom{00}$ & $\phantom{00}719\pm 24\phantom{00}$ & $\phantom{0}1637\pm 40\phantom{00}$ & $\phantom{0}1916\pm 44\phantom{00}$ & $\phantom{0}2442\pm 55\phantom{00}$ & $\phantom{0}3137\pm 58\phantom{00}$ & $\phantom{0}3691\pm 55\phantom{00}$ & \\ 
$10.25\leq \mathcal{M} < 10.50$  & $\phantom{00}351\pm 20\phantom{00}$ & $\phantom{00}437\pm 20\phantom{00}$ & $\phantom{0}1084\pm 32\phantom{00}$ & $\phantom{0}1371\pm 38\phantom{00}$ & $\phantom{0}1768\pm 39\phantom{00}$ & $\phantom{0}2237\pm 48\phantom{00}$ & $\phantom{0}2216\pm 46\phantom{00}$ & $\phantom{00}825\pm 31\phantom{00}$ \\ 
$10.50\leq \mathcal{M} < 10.75$  & $\phantom{00}138\pm 11\phantom{00}$ & $\phantom{00}223\pm 13\phantom{00}$ & $\phantom{00}554\pm 23\phantom{00}$ & $\phantom{00}714\pm 25\phantom{00}$ & $\phantom{0}1050\pm 33\phantom{00}$ & $\phantom{0}1424\pm 37\phantom{00}$ & $\phantom{0}1273\pm 37\phantom{00}$ & $\phantom{00}384\pm 19\phantom{00}$ \\ 
$10.75\leq \mathcal{M} < 11.00$  & $\phantom{000}47\pm \phantom{0}6\phantom{00}$ & $\phantom{000}82\pm \phantom{0}9\phantom{00}$ & $\phantom{00}150\pm 12\phantom{00}$ & $\phantom{00}278\pm 16\phantom{00}$ & $\phantom{00}388\pm 20\phantom{00}$ & $\phantom{00}646\pm 23\phantom{00}$ & $\phantom{00}558\pm 24\phantom{00}$ & $\phantom{00}189\pm 13\phantom{00}$ \\ 
$11.00\leq \mathcal{M} < 11.25$  & $\phantom{0000}7\pm \phantom{0}2\phantom{00}$ & $\phantom{0000}9\pm \phantom{0}3\phantom{00}$ & $\phantom{000}22\pm \phantom{0}5\phantom{00}$ & $\phantom{000}56\pm \phantom{0}8\phantom{00}$ & $\phantom{000}86\pm \phantom{0}9\phantom{00}$ & $\phantom{00}154\pm 12\phantom{00}$ & $\phantom{00}153\pm 13\phantom{00}$ & $\phantom{000}68\pm \phantom{0}9\phantom{00}$ \\ 
$11.25\leq \mathcal{M} < 11.50$  & \phantom{0}-- & $\phantom{0000}2\pm \phantom{0}1\phantom{00}$ & $\phantom{0000}2\pm \phantom{0}1\phantom{00}$ & $\phantom{0000}8\pm \phantom{0}3\phantom{00}$ & \phantom{0}-- & $\phantom{000}14\pm \phantom{0}3\phantom{00}$ & $\phantom{000}14\pm \phantom{0}3\phantom{00}$ & $\phantom{000}21\pm \phantom{0}5\phantom{00}$ \\ 
\hline
\end{tabular}
\tablefoot{Stacked results for star-forming galaxies between redshifts 0.45 and 5.7 (top panel) with the number of stacked optical/near-IR sources in each bin (bottom panel). The first column lists the mass bins, where $\mathcal{M}~\equiv~\log_{10}(M_{\star}/\Msun)$.}
\end{sidewaystable*}

\end{appendix}

\end{document}